\documentclass[pre,tighten]{revtex4}

\usepackage{amssymb,amsmath}

\usepackage{graphicx}

\begin{document}
\title{Diffusion of impurities in a granular gas}
\author{Vicente Garz\'{o}}
\email[E-mail: ]{vicenteg@unex.es}
\address{Departamento de F\'{\i}sica, Universidad de Extremadura, E-06071
Badajoz, Spain}
\author{Jos\'e Mar\'{\i}a Montanero}
\email[E-mail: ]{jmm@unex.es}
\address{Departamento de Electr\'onica e Ingenier\'{\i}a Electromec\'anica,\\
Universidad de Extremadura, E-06071
Badajoz, Spain}

\begin{abstract} 
Diffusion of impurities in a granular gas undergoing homogeneous cooling state is studied. The results are obtained by solving the Boltzmann--Lorentz equation by means of the Chapman--Enskog method. In the first order in the density gradient of impurities, the diffusion coefficient $D$ is determined as the solution of a linear integral equation which is approximately solved by making an expansion in Sonine polynomials. In this paper, we evaluate $D$ up to the second order in the Sonine expansion and get explicit expressions for $D$ in terms of the coefficients of restitution for the impurity--gas and gas--gas collisions as well as the ratios of mass and particle sizes. To check the reliability of the Sonine polynomial solution, analytical results are compared with those obtained from numerical solutions of the Boltzmann equation by means of the direct simulation Monte Carlo (DSMC) method. In the simulations, the diffusion coefficient is measured via the mean square displacement of impurities. The comparison between theory and simulation shows in general an excellent agreement, except for the cases in which the gas particles are much heavier and/or much larger than impurities. In theses cases, the second Sonine approximation to $D$ improves significantly the qualitative predictions made from the first Sonine approximation. A discussion on the convergence of the Sonine polynomial expansion is also carried out.     
\end{abstract}

\draft
\date{\today}
\maketitle

\section{Introduction}
\label{sec1}

The use of hydrodynamic-like type equations to describe the macroscopic behavior of granular media under rapid flow conditions has been widely recognized in the past few years. The main difference from ordinary fluids is the dissipative character of collisions which leads to modifications of the Navier-Stokes equations. In order to gain some insight into the derivation of hydrodynamics from a kinetic theory point of view, an idealized model is usually considered: a system composed by smooth hard spheres with inelastic collisions.  In the low-density regime, the above issue can be addressed by using the Boltzmann kinetic theory conveniently modified to account for inelastic collisions \cite{BDS97}. In this case, assuming the existence of a {\em normal} solution for sufficiently long space and time scales, the Chapman--Enskog method \cite{CC70} can be applied to solve the Boltzmann equation and determine the explicit form of the transport coefficients. This objective has been widely covered in the case of a monocomponent gas \cite{BDKS98,BC01,GM02}, where all the particles have the same mass and size. Nevertheless, a real granular system is generally composed by particles with different mass densities and sizes, which leads to many interesting phenomena observed in nature and experiments such as separation or segregation.

Needless to say, the determination of the transport coefficients of a {\em multicomponent} granular fluid is much more complicated than in the case of a monocomponent system. Several attempts to get those coefficients from the Boltzmann equation began time ago \cite{JM89,Z95,AW98,WA99,AWAL02}, but the technical difficulties for the analysis entailed approximations that limited their range of validity (quasielastic limit). In particular, all these works assume energy equipartition and consequently, the partial temperatures $T_i$ are made equal to the global granular temperature $T$. However, the failure of energy equipartition in granular fluids \cite{GD99,MP99,BT02,MP02,NK02} has also been confirmed by computer simulations \cite{MG02,CH02,DHGD02,PMP02,PCMP03,KT03,WJM03} and even observed in recent experiments \cite{WP02,FM02}. As a consequence, previous studies for granular mixtures based on a single temperature  must be reexamined by theories which take into account the effect of temperature differences on the transport coefficients. For this reason, recently Garz\'o and Dufty \cite{GD02} have carried out a derivation of hydrodynamics for a granular binary mixture at low-density that accounts for nonequipartition of granular energy. Their results provide a description of hydrodynamics in granular mixtures valid a priori over the broadest parameter range and not limited to nearly elastic particles. On the other hand, from a practical point of view, the expressions for the transport coefficients were obtained by considering the leading terms in a Sonine polynomial expansion of the distribution functions. In the case of the shear viscosity coefficient, the first Sonine predictions compare quite well with a numerical solution of the Boltzmann equation in the uniform shear flow state \cite{MG03} obtained from the direct simulation Monte Carlo (DSMC) method \cite{B94}. Exceptions to this agreement are extreme mass or size ratios and strong dissipation.  These discrepancies could be mitigated in part if one considers higher-order terms in the Sonine polynomial expansion.

The evaluation of the transport coefficients for a granular binary mixture beyond the first Sonine approximation is quite a hard task, due mainly to the coupling among the different integral equations associated with the transport coefficients \cite{GD02}.  For this reason, to make some progress into this problem, one needs to study some specific simple situations.  Here, we consider the special case in which one of the components (say, for instance, species 1) is present in {\em tracer} concentration. The tracer problem is more amenable to be treated analytically since the tracer particles (impurities) are enslaved to the granular gas and there are fewer parameters. In this situation, one can assume that the granular gas (excess component) is not affected by the presence of impurities so that its velocity distribution function $f_2$  verifies the (closed) nonlinear Boltzmann equation.  Moreover, the mole fraction of impurities is so small that one can also neglect collisions among tracer particles in their corresponding kinetic equation.  As a consequence, the velocity distribution function $f_1$ of impurities verifies the linear Boltzmann--Lorentz equation.

The goal of this paper is  to determine the diffusion coefficient of impurities immersed in a low-density granular gas undergoing the so-called homogeneous cooling state (HCS). It corresponds to a homogeneous state where the temperature uniformly decreases in time due to dissipation in collisions. 
Diffusion of impurities in a homogeneous gas is perhaps the simplest nonequilibrium problem one can think of. The diffusion coefficient $D$ is obtained by solving the Boltzmann-Lorentz equation by means of the Chapman--Enskog method through the first order in the concentration gradient. As in the elastic case, the coefficient $D$ is expressed in terms of the solution of a linear integral equation that can be solved by making an expansion in Sonine polynomials. An important simplification with respect to the case of arbitrary composition \cite{GD02} is that the diffusion coefficient  satisfies a {\em closed} equation and so is not coupled to the remaining transport coefficients. This simplifies enormously the procedure of getting the different Sonine approximations. Here, we have retained  up to the second Sonine approximation and have determined $D$ in terms of the coefficients of restitution for the impurity--gas and gas--gas collisions, and the parameters of the system (masses and sizes). As for the shear viscosity coefficient \cite{MG03}, kinetic theory predictions are also compared with numerical solutions of the Boltzmann--Lorentz equation by using the DSMC method.  In the simulations, the diffusion coefficient is computed from the mean-square displacement of impurities once a transformation to dimensionless variables allows one to get a stationary diffusion equation.

Some related papers studying diffusion in a homogeneous granular gas have been  published earlier. Thus, Brilliantov and P{\"o}schel \cite{BP00} obtained the self-diffusion coefficient (when impurities and gas particles are mechanically equivalent particles) from a cumulant expansion of the velocity autocorrelation function, while Brey {\em et al.} \cite{BMCG00} evaluated this coefficient in the first Sonine approximation by means of the Chapman--Enskog expansion. In the latter work, the authors also performed Monte Carlo simulations and found a good agreement between theory and simulation. The dynamics of a heavy impurity particle in a gas of much lighter particles has been also studied from the Fokker-Planck equation \cite{BDS99}. The predictions of this (asymptotic) theory have also been confirmed by Monte Carlo and molecular dynamics simulations \cite{BMGD99}. More recently, Dufty {\em et al.} \cite{DBL02} applied the familiar methods of linear response theory to get Green-Kubo and Einstein representations of diffusion coefficient in terms of the velocity and mean-square displacement correlation functions. These correlation functions were also evaluated approximately considering the leading cumulant approximation. This result coincides with the one obtained from the Chapman--Enskog method in the first Sonine approximation \cite{DG01}. The expression for the diffusion coefficient obtained in Ref.\ \cite{DBL02} has been also compared with results from molecular-dynamics 
simulations \cite{LBD02} over a wide range of density and inelasticity, for the particular case of self-diffusion.  The comparison shows that the approximate theory is in good agreement with simulation data up to moderate densities and degrees of inelasticity, although large discrepancies exist at high density and large dissipation. However, these deviations from the Enskog model are primarily due to the effect of an unstable long wavelength shear mode in the system.   
Our work complements and extends the results obtained in previous studies in two aspects: First, our Chapman--Enskog solution is more accurate since incorporates higher order Sonine polynomials, and second, our simulations cover a range of values of the ratios of mass and particles sizes where the effect of energy nonequipartition on diffusion is in general quite important \cite{GD02}. The diffusion phenomenon in granular shear flows has also been widely studied by computer simulations \cite{campbell}, kinetic theory \cite{G02} and real experiments \cite{NHT95}. In this situation, due to the presence of shear flow, the resulting diffusion process is anisotropic and, thus, it must be described by a diffusion tensor.

The plan of the paper is as follows. In Sec.\ \ref{sec2} we describe the problem we are interested in and  analyze the state of the granular gas and impurities in the absence of diffusion.  In particular, we show the breakdown of energy equipartition and illustrate the dependence of the temperature ratio on the parameters of the problem. Section \ref{sec3} deals with the Chapman--Enskog method to solve the Boltzmann--Lorentz equation in the first order of the concentration gradient.  Some technical details of the calculations are given in Appendices \ref{appA} and \ref{appB}. In Sec.\ \ref{sec4} we present the Monte Carlo simulation of the Boltzmann--Lorentz equation and compare the simulation data with the theoretical results obtained in the first and second Sonine approximations. Finally, in Sec.\ \ref{sec5} we close the paper with a brief discussion on the results obtained in this paper.

\section{Granular binary mixture in the homogeneous cooling state}
\label{sec2}

Consider a binary mixture of smooth hard spheres of 
masses $m_{1}$ and $m_{2} $, diameters $\sigma _{1}$ and $\sigma _{2}$, and interparticle coefficients of restitution $\alpha_{11}$, $\alpha_{22}$, and 
$ \alpha_{12}=\alpha_{21}$. Here, $\alpha_{ij}$ is the coefficient of restitution for collisions between particles of species $i$ and $j$. In the low-density regime, the 
distribution functions $f_{i}({\bf r},{\bf v};t)$ $
(i=1,2)$ for the two species are determined from the set of nonlinear 
Boltzmann equations \cite{BDS97}
\begin{equation} 
\left( \partial _{t}+{\bf v}_{1}\cdot \nabla \right) 
f_{i}=\sum_{j}J_{ij}\left[ {\bf v}_{1}|f_{i}(t),f_{j}(t)\right] \;,
\label{2.1} 
\end{equation} 
where  the Boltzmann collision operator $J_{ij}\left[ {\bf v}_{1}|f_{i},f_{j}\right] $ is 
\begin{eqnarray} 
J_{ij}\left[ {\bf v}_{1}|f_{i},f_{j}\right] &=&\sigma _{ij}^{2}\int d{\bf v} 
_{2}\int d\widehat{\boldsymbol {\sigma }}\,\Theta (\widehat{{\boldsymbol {\sigma }}} 
\cdot {\bf g}_{12})(\widehat{\boldsymbol {\sigma }}\cdot {\bf g}_{12})  \nonumber 
\\ &&\times \left[ \alpha _{ij}^{-2}f_{i}({\bf r},{\bf v}_{1}^{\prime 
},t)f_{j}( {\bf r},{\bf v}_{2}^{\prime },t)-f_{i}({\bf r},{\bf v}
_{1},t)f_{j}({\bf r}, {\bf v}_{2},t)\right] \;.  \label{2.2} 
\end{eqnarray} 
Here, $\sigma _{ij}=\left( \sigma _{i}+\sigma _{j}\right) /2$, $\widehat{
\boldsymbol {\sigma}}$ is a unit vector along their line of centers, $\Theta $ is 
the Heaviside step function, and ${\bf g}_{12}={\bf v}_{1}-{\bf v}_{2}$. The 
primes on the velocities denote the initial values $\{{\bf v}_{1}^{\prime},  
{\bf v}_{2}^{\prime }\}$ that lead to $\{{\bf v}_{1},{\bf v}_{2}\}$ 
following a binary collision:  
\begin{equation} 
{\bf v}_{1}^{\prime }={\bf v}_{1}-\mu _{ji}\left( 1+\alpha _{ij}^{-1}\right) 
(\widehat{{\boldsymbol {\sigma }}}\cdot {\bf g}_{12})\widehat{{\boldsymbol {\sigma }}} 
,\quad {\bf v}_{2}^{\prime }={\bf v}_{2}+\mu _{ij}\left( 1+\alpha 
_{ij}^{-1}\right) (\widehat{{\boldsymbol {\sigma }}}\cdot {\bf g}_{12})\widehat{ 
\boldsymbol {\sigma}} , \label{2.3} 
\end{equation} 
where $\mu _{ij}=m_{i}/\left( m_{i}+m_{j}\right)$.

The relevant hydrodynamic fields are the number densities $n_{i}$, 
the flow velocity ${\bf u}$, and the ``granular'' temperature $T$. 
They are defined in terms of moments of the distributions $f_{i}$ as  
\begin{equation} 
n_{i}=\int d{\bf v}f_{i}({\bf v})\;,\quad \rho {\bf u}=\sum_{i}\int 
d {\bf v}m_{i}{\bf v}f_{i}({\bf v})\;,  \label{2.4} 
\end{equation} 
\begin{equation} 
nT=p=\sum_{i}\int d{\bf v}\frac{m_{i}}{3}V^{2}f_{i}({\bf v})\;, 
\label{2.5} 
\end{equation} 
where $n=n_{1}+n_{2}$ is the total number density, $\rho 
=m_{1}n_{1}+m_{2}n_{2}$ is the total mass density, $p$ is the 
hydrostatic pressure, and ${\bf V}={\bf v}-{\bf u}$ is the peculiar velocity. Furthermore, it is convenient to introduce the 
kinetic temperatures $T_i$ for each species, which measure their mean kinetic energies. They are defined as 
\begin{equation}
\label{2.6}
\frac{3}{2}n_iT_i=\int d{\bf v}\frac{m_{i}}{2}V^{2}f_{i},
\end{equation}
so that the granular temperature $T$ is 
\begin{equation}
\label{2.6bis}
T=\sum_i x_iT_i, 
\end{equation}
where $x_i=n_i/n$ is the mole fraction of species $i$.

The collision operators conserve the particle number of each species and the 
total momentum but the total energy is not conserved:
\begin{equation}
\label{2.7}
\int d{\bf v}J_{ij}[{\bf v}|f_i,f_j]=0,\quad 
\sum_{i,j}\int d{\bf v}m_i{\bf v}J_{ij}[{\bf v}|f_i,f_j]=0,
\end{equation}
\begin{equation}
\label{2.8}
3nT\zeta=-\sum_{i,j}\int d{\bf v}m_iV^2J_{ij}[{\bf 
v}|f_i,f_j],
\end{equation}
where $\zeta$ is identified as the {\em cooling rate} due to inelastic 
collisions among all species. At a kinetic level, it is convenient to characterize energy 
transfer in terms of the {\em cooling rates} associated with the partial temperatures $T_i$. They are defined as 
\begin{equation}
\label{2.9}
3n_iT_i\zeta_i=-\sum_{j}\int d{\bf v}m_iV^2J_{ij}[{\bf v}|f_i,f_j].
\end{equation}
The total cooling rate $\zeta$ can be expressed in terms of the partial cooling rates $\zeta_i$ as 
\begin{equation}
\label{2.10}
\zeta=T^{-1}\sum_i x_iT_i\zeta_i.
\end{equation}

Because of the complexity embodied in the general description of a binary mixture, here we consider the special case in which the mole fraction of one of the components (say, for instance, 1) is negligible ($x_1 \ll 1$). We are interested in studying the diffusion of impurities moving in a background granular gas undergoing homogeneous cooling state (HCS). In the tracer limit, one expects that the state of the granular gas (solvent) is not disturbed by the presence of impurities (solute) and so in all of the following it is assumed that the gas is in its HCS. In addition, collisions among impurities themselves can be neglected versus the impurity--gas collisions. Under these conditions, the velocity distribution function $f_2$ of the gas verifies a (closed) Boltzmann equation and the velocity distribution function $f_1$ of impurities obeys a (linear) Boltzmann--Lorentz equation. Let us start by describing the state of the system in the absence of diffusion.  

\subsection{Granular gas}

As said above, the gas is in the HCS. This state corresponds to a homogeneous solution of the nonlinear Boltzmann equation, $f_2(v,t)$,  in which all the time dependence occurs through the temperature of the gas $T_2(t)\simeq T(t)$. In this case, the time derivative of $f_2$ can be represented more usefully as
\begin{equation}
\label{2.11}
\partial_t f_2=-\zeta_2 T\partial_T f_2=\frac{1}{2}\zeta_2\frac{\partial}{\partial {\bf v}} \cdot \left({\bf v}f_2\right),
\end{equation}
where use has been made of the balance equation for the temperature
\begin{equation}
\label{2.11bis}
T^{-1}\partial_t T=-\zeta_2 ,
\end{equation}
with 
\begin{equation}
\label{2.12bis}
\zeta_2=-\frac{1}{3n_2T}\int d{\bf v}m_2v^2J_{22}[f_2,f_2].
\end{equation}
The Boltzmann equation can be written as 
\begin{equation}
\label{2.13}
\frac{1}{2}\zeta_2\frac{\partial}{\partial {\bf v}} \cdot \left({\bf v}f_2\right)=J_{22}[f_2,f_2].
\end{equation}
The second equality in (\ref{2.11}) follows from dimensional analysis which requires that the temperature dependence of $f_2$ occurs only through the temperature $T(t)$. Consequently, $f_2(v,t)$ has the form
\begin{equation}
\label{2.12}
f_2(v,t)=n_2\pi^{-3/2}v_0^{-3/2}(t)\Phi_2(v/v_0(t)),
\end{equation}
where $v_0(t)=\sqrt{2T/m_2}$ is the thermal velocity of the gas. So far, the exact form of $\Phi_2$ has not been found, although a good approximation for thermal velocities can be obtained from an expansion in Sonine polynomials. In the leading order, $\Phi_2$ is given by 
\begin{equation}
\label{2.14}
\Phi_2(v^*)\to  \left[ 1+\frac{c_2}{4}
\left(v^{*4}-5v^{*2}+\frac{15}{4}\right)\right]e^{-v^{*2}}.
\end{equation} 
The coefficient $c_2$ can be obtained by substituting first Eqs.\ (\ref{2.12}) and (\ref{2.14}) into the Boltzmann equation (\ref{2.13}), multiplying that equation by $v^4$ and then integrating over the velocity. When only linear terms in $c_2$ are retained,  the estimated value of $c_2$ is \cite{NE98}
\begin{equation}
\label{2.15}
c_2(\alpha_{22})=\frac{32(1-\alpha_{22})(1-2\alpha_{22}^2)}
{81-17\alpha_{22}+30\alpha_{22}^2(1-\alpha_{22})}.
\end{equation} 
Additionally, the cooling rate $\zeta_2$ can also be determined from the Sonine approximation (\ref{2.14}). The result is \cite{NE98}
\begin{equation}
\label{2.15bis}
\zeta_2=\frac{2}{3}\sqrt{2\pi}n_2\sigma_{2}^2v_0(1-\alpha_{22}^2)\left(1+\frac{3}{32}c_2\right).
\end{equation} 
Estimate (\ref{2.15}) presents quite a good agreement with Monte Carlo simulations of the Boltzmann equation \cite{BMC96,MS00}.

\subsection{Impurities}

In the absence of diffusion, impurities are also in HCS and so its velocity distribution function $f_1(v,t)$ satisfies the Boltzmann--Lorentz equation 
\begin{equation}
\label{2.16}
\frac{1}{2}\zeta_1\frac{\partial}{\partial {\bf v}} \cdot \left({\bf v}f_1\right)=J_{12}[f_1,f_2].
\end{equation}
The time evolution of $T_1(t)$ is 
\begin{equation}
\label{2.16bis}
T_1^{-1}\partial_t T_1=-\zeta_1,
\end{equation}
where 
\begin{equation}
\label{2.16.1bis}
\zeta_1=-\frac{1}{3n_1T_1}\int d{\bf v}m_1v^2J_{12}[f_1,f_2].
\end{equation}
From Eqs.\ (\ref{2.11bis}) and (\ref{2.16bis}) one easily gets the time evolution of the temperature ratio $\gamma\equiv T_1/T$:
\begin{equation}
\label{2.16.1}
\gamma^{-1}\partial_t \gamma=\zeta_2-\zeta_1 .
\end{equation}
The fact that $f_1$ depends on time only through $T(t)$ necessarily implies  that the temperature 
ratio $\gamma$ must be independent on time, and so Eq.\ (\ref{2.16.1}) gives the HCS condition $\zeta_1(t)=\zeta_2(t)=\zeta(t)$. However, although both components have a common cooling rate, their partial temperatures are different. In other words, the impurity {\em equilibrates} to a common HCS with different temperatures for the impurity and gas particles. This implies a breakdown of the energy equipartition. The violation of energy equipartition in multicomponent granular systems has been even observed in real experiments of vibrated mixtures in two \cite{WP02} and three \cite{FM02} dimensions.

\begin{figure}
\includegraphics[width=0.4 \columnwidth]{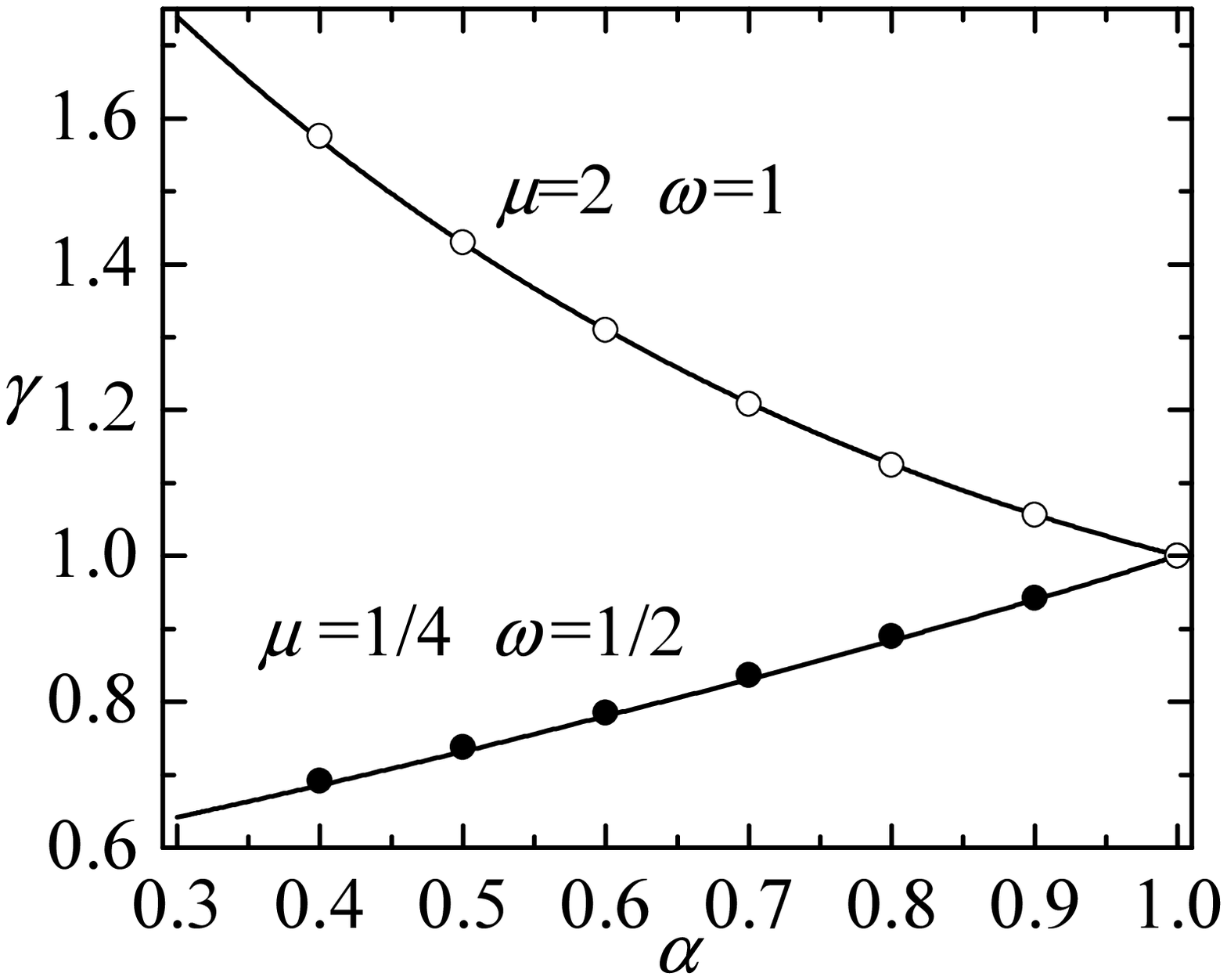}
\caption{Temperature ratio $\gamma=T_1/T$ versus the coefficient of restitution  $\alpha\equiv\alpha_{12}=\alpha_{22}$ for $\mu\equiv m_1/m_2=2$ and $\omega\equiv \sigma_1/\sigma_2=1$ (open circles) and $\mu=1/4$ and $\omega=1/2$ (filled circles). The lines are the theoretical results and the symbols refer to the numerical results obtained from  the DSMC method.
\label{fig1}}
\end{figure}

As for the gas, dimensional analysis requires that the solution to Eq.\ (\ref{2.16}) is of the form
\begin{equation}
\label{2.17}
f_1(v,t)=n_1\pi^{-3/2}v_0^{-3/2}(t)\Phi_1(v/v_0(t)).
\end{equation}
The determination of $f_1$ to leading order in the Sonine expansion has been analyzed elsewhere for arbitrary composition \cite{GD99} and only the main results are quoted here. In the first Sonine approximation, the distribution $\Phi_1$ is given by  
\begin{equation}
\label{2.18}
\Phi_1(v^*)\to  \theta^{3/2}\left[ 1+\frac{c_1}{4}
\left(\theta^2v^{*4}-5\theta v^{*2}+\frac{15}{4}\right)\right]e^{-\theta v^{*2}},
\end{equation} 
where $\theta=m_1T/m_2T_1$ is the mean-square velocity of the gas particles relative to that of  the tracer particles.  The coefficient $c_1$ in Eq.\ (\ref{2.18}) can be determined by substitution of Eqs.\ (\ref{2.12}) and (\ref{2.17}) into Eq.\ (\ref{2.16}) and retaining all terms linear in $c_1$ and $c_2$ for the leading polynomial approximations (\ref{2.14}) and (\ref{2.18}). Once the coefficient $c_1$ is known, one can estimate the temperature ratio $\gamma\equiv T_1/T$ from the constraint $\zeta_1=\zeta_2$. The solution to this equation gives $\gamma$ as a function of the mass ratio $\mu\equiv m_1/m_2$, the size ratio $\omega\equiv \sigma_1/\sigma_2$ and the coefficients of restitution $\alpha_{22}$ and $\alpha_{12}$. The explicit expressions of the partial cooling rate $\zeta_1$ and the coefficient $c_1$ are displayed in Appendix \ref{appC}.

Except for some limiting cases (elastic case and mechanically equivalent particles), our results yield $\gamma\neq 1$, and so the total energy is not equally distributed between both species. The lack of energy equipartition has dramatic consequences in the large impurity/gas mass ratio since there is a peculiar ``phase transition'' for which the diffusion coefficient is normal in one phase and grows without bound in the other \cite{SD01}.  To illustrate the violation of equipartition theorem, in Fig.\ \ref{fig1} we plot the temperature ratio versus the coefficient of restitution $\alpha$ for two different cases: $\mu=1/4$, $\omega=1/2$, and $\mu=2$, $\omega=1$. For the sake of simplicity, we have taken a common coefficient of restitution, i.e., $\alpha\equiv \alpha_{12}=\alpha_{22}$. We also include the simulation data obtained via a numerical solution of the Boltzmann equation by means of the direct simulation Monte Carlo (DSMC) method \cite{B94}. It is apparent the excellent agreement found between the theory and simulation, showing the accuracy of the approximations (\ref{2.14}) and (\ref{2.18}) to estimate the temperature ratio. We also observe that the temperature of the impurity is larger than that of the gas when the impurity is heavier than the grains of the gas. In addition, the deviations from the energy equipartition increase as the mechanical differences between impurities and particles of the gas increase. As we will show later, in general the effect of the temperature differences on the diffusion coefficient is quite important.

\section{Diffusion coefficient}
\label{sec3}

We want to determine the diffusion coefficient of tracer particles immersed in a granular gas in HCS. The diffusion process is induced by a weak concentration gradient $\nabla x_1$, which is the only gradient present in the system.  Under these conditions, the kinetic equation for $f_1$ reads 
\begin{equation}
\label{3.1}
\left( \partial _{t}+{\bf v}_{1}\cdot \nabla \right) f_{1}=J_{12}\left[ {\bf v}_{1}|f_{1},f_{2}\right] \;.
\end{equation} 
Impurities may freely exchange momentum and energy with the particles of the granular gas, and, therefore, these are not invariants of the Boltzmann--Lorentz collision operator $J_{12}[f_1,f_2]$. Only the number density of impurities is conserved: 
\begin{equation}
\label{3.2}
\partial_t n_1+\frac{\nabla \cdot {\bf j}_1}{m_1}=0,
\end{equation}
where the flux  of tracer particles ${\bf j}_1$ is defined as 
\begin{equation}
\label{3.3}
{\bf j}_1=m_1\int d{\bf v} {\bf v}f_1({\bf v}).
\end{equation}

The conservation equation (\ref{3.2}) becomes a closed hydrodynamic equation for $n_1$ once ${\bf j}_1$ is expressed as a functional of the fields $n_1$ and $T$. Our aim is to get the mass flux in the first order in $\nabla x_1$ by applying the Chapman--Enskog method \cite{CC70}. The Chapman--Enskog method assumes the existence of a {\em normal} solution in which all the space and time dependence of $f_1$ occurs through the hydrodynamic fields. In the tracer diffusion problem, this means that 
\begin{equation}
\label{3.4}
f_1({\bf r}, {\bf v};t)=f_1[{\bf v}|x_1({\bf r}, t), T(t)],
\end{equation}
where it has been assumed that $f_2$ also adopts the normal form (\ref{2.12}). The Chapman--Enskog procedure generates the normal solution explicitly by means of an expansion in gradients of the fields  
\begin{equation}
\label{3.5}
f_1=f_1^{(0)}+\epsilon f_1^{(1)}+\epsilon^2 f_1^{(2)}+\cdots,
\end{equation}
where $\epsilon$ is a formal parameter measuring the nonuniformity of the system. Given that in our problem the only inhomogeneity is in the mole fraction $x_1$, each factor $\epsilon$ corresponds to an implicit $\nabla x_1$ factor. The time derivative is also expanded as $\partial_t=\partial_t^{(0)}+\epsilon \partial_t^{(1)}+\cdots$, where 
\begin{equation}
\label{3.6}
\partial_t^{(0)}x_1=0,\quad \partial_t^{(0)}T=-T\zeta,
\end{equation}
\begin{equation}
\label{3.7}
\partial_t^{(k)}x_1=-\frac{\nabla \cdot {\bf j}_1^{(k-1)}}{m_1n_2},\quad \partial_t^{(k)}T=0,\quad k \geq 1,
\end{equation}
with 
\begin{equation}
\label{3.8}
{\bf j}_1^{(k)}=m_1\int d{\bf v} {\bf v}f_1^{(k)}.
\end{equation}
Upon deriving Eqs.\ (\ref{3.6}) and (\ref{3.7}), use has been made of the balance equations (\ref{2.11bis}) and (\ref{3.2}) and the constraint $\zeta_1^{(0)}=\zeta_2=\zeta$. Here, $\zeta_1^{(0)}$ is given by Eq.\ (\ref{2.16.1bis}) with $f_1\to f_1^{(0)}$.

The zeroth-order approximation $f_1^{(0)}$ is the solution of (\ref{2.16}), whose approximate form is given by Eqs.\ (\ref{2.17}) and (\ref{2.18}) but taking into account now the local dependence on the mole fraction $x_1$. Since $f_1^{(0)}$ is isotropic, it follows that the flux of impurities vanishes at this order, i.e., ${\bf j}_1^{(0)}={\bf 0}$, and so $\partial_t^{(1)}x_1=0$. To first order in $\epsilon$, one has the kinetic equation
\begin{eqnarray}
\label{3.9}
\partial _{t}^{(0)}f_1^{(1)}+J_{12}[f_{1}^{(1)},f_{2}]&=&-\left(\partial_t^{(1)}+{\bf v}_1\cdot \nabla \right)f_1^{(0)}\nonumber\\
&=& -\left(\frac{\partial}{\partial x_1}f_1^{(0)}\right){\bf v}_1\cdot \nabla x_1.
\end{eqnarray} 
The second equality follows from the balance equations (\ref{3.7}) and the space dependence of $f_1^{(0)}$ through $x_1$. The solution to Eq.\ (\ref{3.9}) is proportional to $\nabla x_1$, namely, it has the form
\begin{equation} 
\label{3.10}
f_1^{(1)}={\boldsymbol {\cal A}}\cdot \nabla x_1.
\end{equation}
The coefficient ${\boldsymbol {\cal A}}$ is a function of the velocity and the hydrodynamic fields $x_1$ and $T$. According to (\ref{3.6}), the time derivative $\partial_t^{(0)}$ acting on ${\boldsymbol {\cal A}}$ can be evaluated by the replacement $\partial_t^{(0)}\to -\zeta T\partial_T$ where $\zeta$ is given by (\ref{2.15bis}). Thus, substitution of (\ref{3.10}) into (\ref{3.9}) yields
\begin{equation}
\label{3.11}
-\zeta T\partial_T{\boldsymbol {\cal A}}-J_{12}[{\boldsymbol {\cal A}},f_2]=-\left(\frac{\partial}{\partial x_1}f_1^{(0)}\right){\bf v}.
\end{equation}
To first order, the mass flux has the structure \cite{M89}
\begin{equation}
\label{3.12}
{\bf j}_1^{(1)}=-m_1D \nabla x_1,
\end{equation}
where $D$ is the diffusion coefficient.  Use of Eq.\ (\ref{3.10}) into the definition (\ref{3.8}) (for $k=1$) allows one to identify the coefficient $D$. It is given by  
\begin{equation}
\label{3.13}
D=-\frac{1}{3}\int d{\bf v}{\bf v} \cdot {\boldsymbol {\cal A}}.
\end{equation}
 
For practical purposes, the linear integral equation (\ref{3.11}) can be solved by using a Sonine polynomial expansion. Our goal here is to determine the diffusion coefficient up to the second 
Sonine approximation. In this case, the quantity ${\boldsymbol {\cal A}}$ is approximated by 
\begin{equation}
\label{3.14}
{\boldsymbol {\cal A}}\to f_{1,M}\left[a_{1}{\bf v}+a_{2}{\bf S}_1({\bf v})\right], 
\end{equation}
where $f_{1,M}$ is a Maxwellian distribution at the temperature $T_1$ of the impurities, i.e.
\begin{equation}
\label{3.15}
f_{1,M}({\bf v})=n_1\left(\frac{m_1}{2\pi T_1}\right)^{3/2}\exp\left(-\frac{m_1v^2}{2T_1}\right),
\end{equation}
and ${\bf S}_1({\bf v})$ is the polynomial 
\begin{equation}
\label{3.16}
{\bf S}_1({\bf v})=\left(\frac{1}{2}m_1v^2-\frac{5}{2}T_1\right){\bf v}.
\end{equation}
The coefficients $a_{1}$ and $a_{2}$ are defined as 
\begin{equation}
\label{3.17}
a_{1}=\frac{m_1}{3n_1T_1}\int d{\bf v}{\bf v} \cdot {\boldsymbol {\cal A}}=-\frac{m_1D}{n_1T_1}, 
\end{equation}
\begin{equation}
\label{3.18}
a_{2}=\frac{2}{15}\frac{m_1}{n_1T_1^3}\int d{\bf v}{\bf S}_1({\bf v}) 
\cdot {\boldsymbol {\cal A}}.
\end{equation}
These coefficients are determined by substitution of Eq.\ (\ref{3.14}) into the integral equation (\ref{3.11}). The details are carried out in Appendix \ref{appA}. The second Sonine approximation $D[2]$  to the diffusion coefficient can be written as 
\begin{equation}
\label{3.19}
D[2]=D[1]\frac{(\nu_a^*-\case{1}{2}\zeta^*)
[\nu_d^*-\case{3}{2}\zeta^*-(c_1/2\gamma)\nu_b^*]}
{(\nu_a^*-\case{1}{2}\zeta^*)(\nu_d^*-\case{3}{2}\zeta^*)-
\nu_b^*[\nu_c^*-(\zeta^*/\gamma)]}\equiv D[1] \Lambda ,
\end{equation}
where the function $\Lambda$ can be easily identified from the 
second equality and $D[1]$ refers to the first Sonine approximation 
to the diffusion coefficient. Its expression is  
\begin{equation}
\label{3.20}
D[1]=\frac{n_2T}{m_1\nu_0}\frac{\gamma}{\nu_a^*-\case{1}{2}\zeta^*}.
\end{equation} 
Here, $\nu_0=n_2\sigma_2^2v_0$,
\begin{equation}
\label{3.21}
\zeta^*=\frac{\zeta}{\nu_0}=\frac{2}{3}\sqrt{2\pi}(1-\alpha_{22}^2)\left(1+\frac{3}{32}c_2\right),
\end{equation}
and the quantities $\nu_a^*$, $\nu_b^*$, $\nu_c^*$, and $\nu_d^*$ are given in Appendix \ref{appB}. Expression  (\ref{3.20}) coincides with the one previously obtained \cite{DG01,DBL02} from a Green-Kubo formula in the leading order in a cumulant expansion.

\begin{figure}
\includegraphics[width=0.4 \columnwidth]{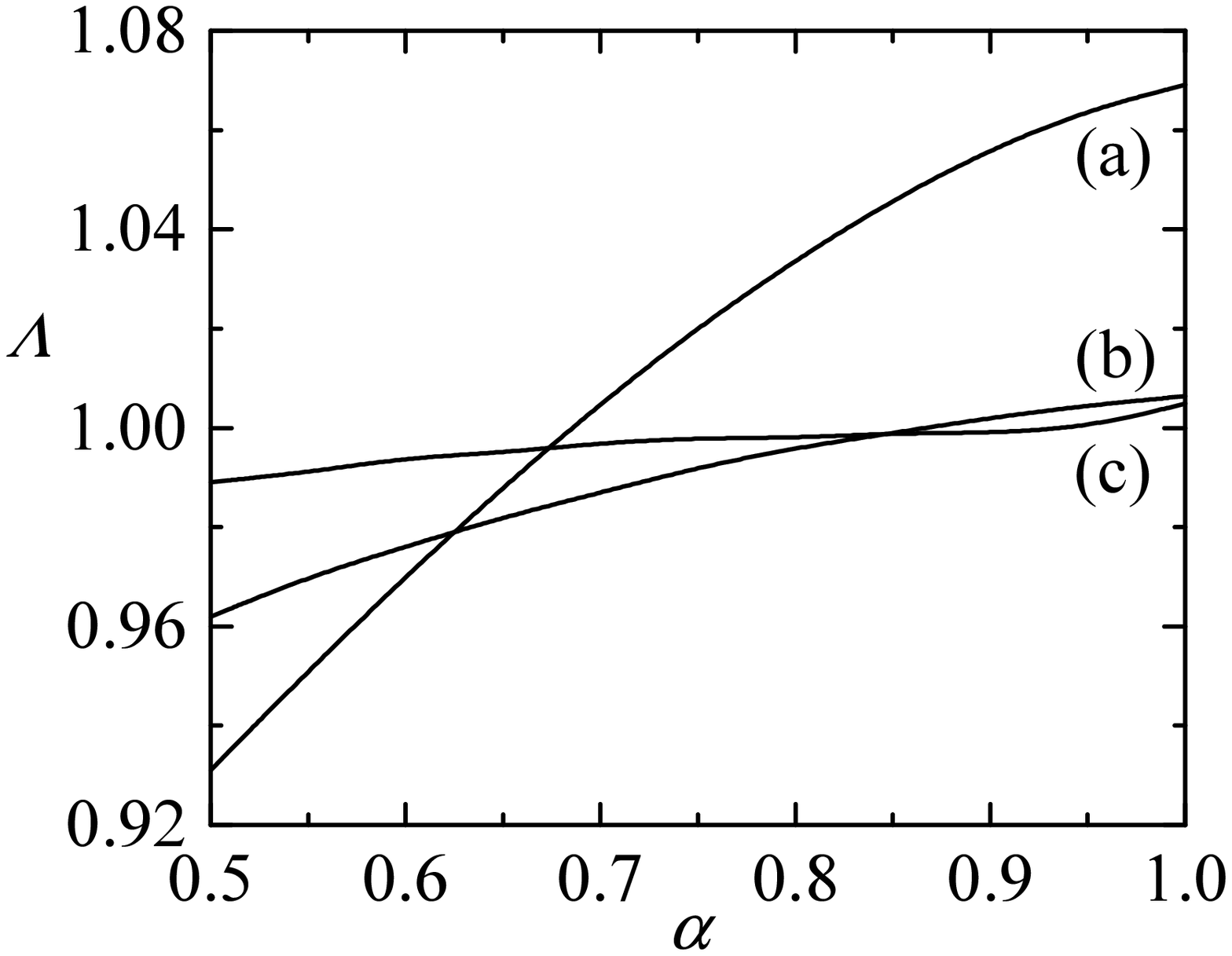}
\caption{Plot of the function $\Lambda\equiv D[2]/D[1]$ versus the coefficient of restitution $\alpha$ for three binary mixtures: (a) $\mu=1/8$ and $\omega=1/2$; (b) $\mu=125/64$ and $\omega=5/4$; and (c) $\mu=8$ and $\omega=2$. 
\label{fig2}}
\end{figure}

In general, $D[1]$ and $D[2]$ present a complex dependence on the coefficients of restitution and the ratios of mass and sizes. Before analyzing this dependence, it is instructive to consider some special cases. In the elastic limit ($\alpha_{22}=\alpha_{12}=1$), one has $c_1=c_2=0$, $\gamma=1$, and the first $D_0[1]$ and second $D_0[2]$ Sonine approximations to the diffusion coefficient become
\begin{equation}
\label{3.22}
D_0[1]=\frac{3}{8}\frac{\sqrt{(m_1+m_2)T}}{\sqrt{2\pi}\sigma_{12}^2m_1},
\end{equation}
\begin{equation}
\label{3.23}
D_0[2]=D_0[1]\frac{30\mu^2+16\mu+13}{30\mu^2+16\mu+12}.
\end{equation}
Equations (\ref{3.22}) and (\ref{3.23}) coincide with the ones derived by Mason \cite{M54} for a gas-mixture of elastic hard spheres. Moreover,  in the case of mechanically equivalent particles ($m_1=m_2$, $\sigma_1=\sigma_2$, $\alpha_{22}=\alpha_{12}=\alpha$), Eq.\ (\ref{3.20}) reduces to the one derived in the self-diffusion problem \cite{BMCG00}.

The function $\Lambda\equiv D[2]/D[1]$ is plotted in Fig.\ \ref{fig2} versus the coefficient of restitution $\alpha$ for three different values of the mass ratio $\mu$ and the size ratio $\omega$. We consider a binary mixture where the mass density of the impurity is equal to that of a bath particle and so, $\mu=\omega^3$. As in the elastic case, we see that $\Lambda$ clearly differs from 1 when impurities are much lighter than the particles of the gas. This means that the widely used first Sonine approximation is not sufficiently accurate for this range of values of mass and diameter ratios. On the other hand, the convergence of the Sonine polynomial expansion improves when increasing the mass and size ratios, and the first Sonine approximation is expected to be quite close to the exact value of the diffusion coefficient. These conclusions are qualitatively similar to those found in the case of elastic  hard-sphere mixtures \cite{MC84}.  

To close this section, let us write the diffusion equation.  In the hydrodynamic regime (where the normal solution to the Boltzmann--Lorentz equation holds), the diffusion coefficient $D(t)$ depends on time {\em only} through its dependence on the temperature $T(t)$. According to Eqs.\ (\ref{3.19}) and (\ref{3.20}), $D(t)\propto \sqrt{T(t)}$. This time dependence can be eliminated by introducing appropriate dimensionless variables. A convenient set of dimensionless time and space variables is given by 
\begin{equation}
\label{3.24}
\tau=\int_{0}^{t} dt'\; \nu_0(t'), \quad {\boldsymbol {\ell}}=\frac{\nu_0(t)}{v_0(t)}{\bf r}.
\end{equation} 
The dimensionless time scale $\tau$ is the integral of the average collision frequency and thus is a measure of the average number of collisions per gas particle in the time interval between 0 and $t$. The unit length $v_0(t)/\nu_0(t)$ introduced in the second equality of (\ref{3.24}) is proportional to the time-independent mean free path of gas particles. In terms of the above variables, the diffusion equation (\ref{3.2}) becomes
\begin{equation}
\label{3.25}
\frac{\partial x_1}{\partial \tau}=D^*\nabla_{\ell}^2x_1,
\end{equation}
where $\nabla_{\ell}^2$ is the Laplace operator in ${\boldsymbol {\ell}}$ space and 
\begin{equation}
\label{3.26} 
D^*=\frac{\nu_0(t)D(t)}{n_2v_0^2(t)}.
\end{equation} 
While $D(t)\propto \sqrt{T(t)}$, the reduced diffusion coefficient $D^*$ is time-independent. This is closely related with the validity of a hydrodynamic description of the system for times large compared with the mean free time. In this context, Eq.\ (\ref{3.25}) is a diffusion equation with a constant diffusion coefficient $D^*$. It follows that the mean square deviation of the {\em position} ${\boldsymbol {\ell}}$ of impurities after a {\em time} interval $\tau$ is \cite{M89}
\begin{equation}
\label{3.27}
\langle  |{\boldsymbol {\ell}}(\tau)-{\boldsymbol {\ell}}(0)|^2 \rangle =6 D^* \tau.
\end{equation}
Restoring the dimensions to the average position and time, Eq.\ (\ref{3.27}) can be rewritten as 
\begin{equation}
\label{3.28}
\frac{\partial}{\partial t} \langle  |{\bf r}(t)-{\bf r}(0)|^2 \rangle=\frac{6D(t)}{n_2}.
\end{equation} 
Equation (\ref{3.28}) is the Einstein form, relating the diffusion coefficient to the mean-square displacement. This relationship will be used later to measure the diffusion coefficient by means of the DSMC method. 

\begin{figure}
\includegraphics[width=0.4 \columnwidth]{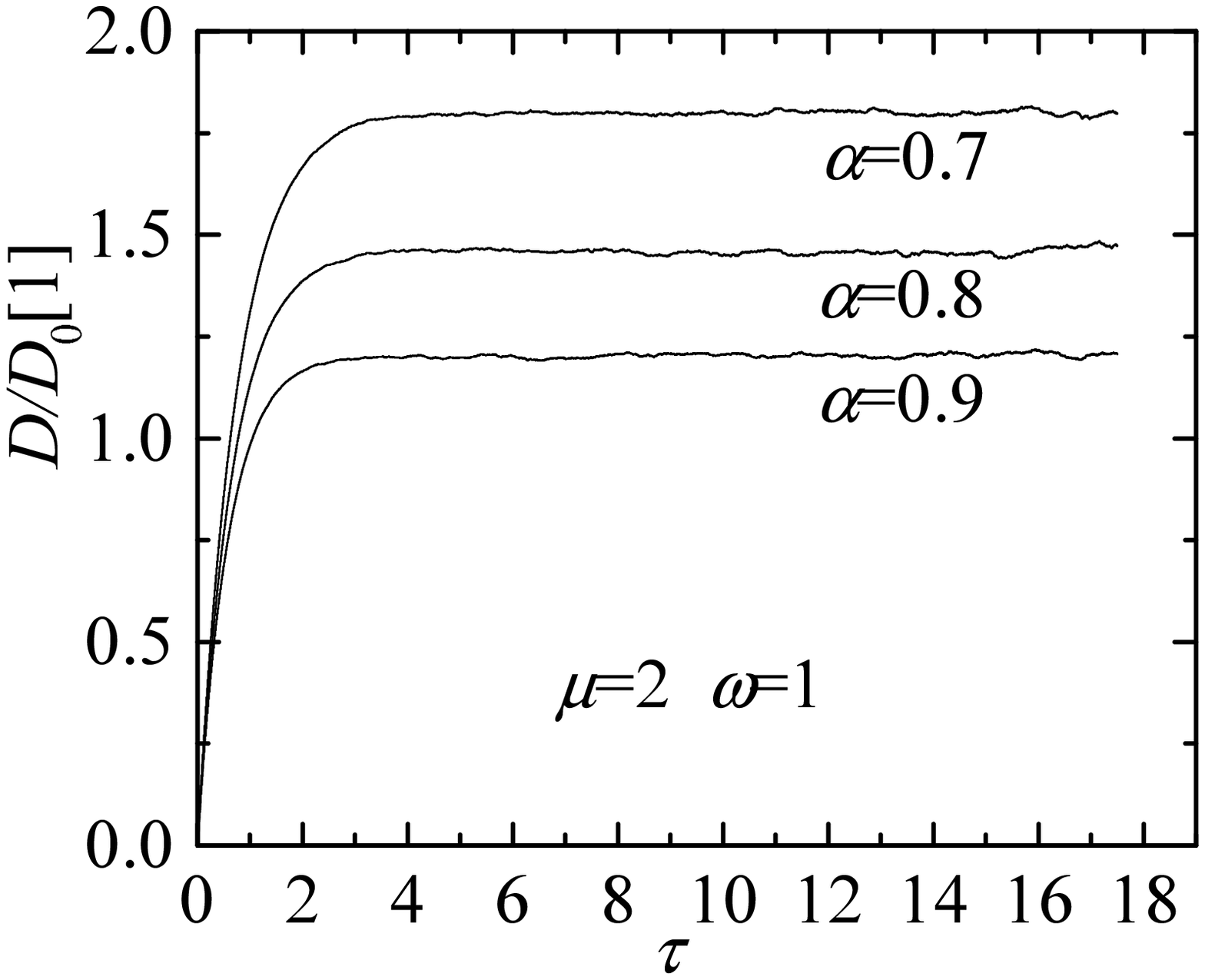}
\caption{Plot of $D/D_0[1]$ as a function of the dimensionless time $\tau$ for $\mu=2$, $\omega=1$ and $\alpha=0.7$, 0.8 and 0.9. Here, $D_0[1]$ refers to the first Sonine approximation to the elastic value of the diffusion coefficient. 
\label{fig3}}
\end{figure}

\section{Comparison between theory and Monte Carlo simulations}
\label{sec4}

The adaptation of DSMC method to analyze binary granular mixtures has been described in detail in previous works (see, for instance, Ref.\ \cite{MG02}), so that here we only shall mention the aspects related to the specific problem of diffusion of impurities in a granular gas under HCS. The main difference between the procedure used here and the one described in Ref.\ \cite{MG02} is its restriction to the tracer limit ($x_1\to 0$). Because of this, during our simulations collisions 1--1  are not considered, and when a collision 1--2 takes place, the postcollisional velocity obtained from the scattering rule is only assigned to the particle 1. According to this scheme, the numbers of particles $N_i$ have simply a statistical meaning, and hence they can be chosen arbitrarily.

In the course of the simulation one can distinguish two stages. In the first one, the system (impurities and gas particles) evolves from the equilibrium initial state to the HCS. In the second stage, the system is assumed to be in the HCS, and then the kinetic temperatures $T_i(t)$ and the diffusion coefficient $D(t)$ are measured. The latter is obtained from the mean square deviation of the position of impurities [Eq.\ (\ref{3.28})], i.e.,
\begin{equation}
\label{3.28b}
D(t)=\frac{n_2}{6\Delta t}\left(\langle  |{\bf r}_i(t+\Delta t)-{\bf r}_i(0)|^2 \rangle -\langle  |{\bf r}_i(t)-{\bf r}_i(0)|^2 \rangle\right).
\end{equation} 
Here, $|{\bf r}_i(t)-{\bf r}_i(0)|$ is the distance traveled by the impurity $i$ from $t=0$ until time $t$, $t=0$ being the beginning of the second stage. In addition, $\langle  \cdots \rangle$ denotes the average over the $N_1$ impurities and $\Delta t$ is the time step. 
In our simulations we have typically taken $10^5$ particles of the granular gas, 
$5\times 10^4$ impurities, and a time step $\Delta t=3\times 10^{-4} \lambda/v_0$, 
where $\lambda=(\sqrt{2}\pi n_2\sigma_{2})^{-1}$ is the mean free path for collisions among 
granular gas particles. The results are averaged  over a number of ${\cal N}=5$ replicas.

If the hydrodynamic description (or normal solution in the context of the Chapman--Enskog method) applies one expects that, after a transient regime in which each gas particle has collided about five times, the reduced diffusion coefficient 
\begin{equation}
\label{3.28c}
\frac{D(t)}{D_0[1](t)}=\frac{16\sqrt{\pi}}{3}\left(\frac{\sigma_{12}}{\sigma_2}\right)^2\frac{\mu_{12}}
{\sqrt{\mu_{21}}}D^*
\end{equation} 
achieves a time-independent plateau. This is illustrated in Fig.\ \ref{fig3} for a system with $\mu=2$, $\omega=1$ and three values of $\alpha: \alpha=0.7$, 0.8 and 0.9. The dimensionless time scale $\tau$ is defined in Eq.\ (\ref{3.24}) and measures the average number of collisions per gas particle, which is of the same order as the corresponding value for impurities. We observe that after several collisions, $D/D_0[1]$ reaches a stationary value whose time average is the simulation result for the (reduced) diffusion coefficient.

The steady state values of $D/D_0[1]$ obtained from simulation data (which are calculated by averaging over a time period) can be compared with the theoretical predictions  $D[2]/D_0[1]$, Eq.\ (\ref{3.19}), and  $D[1]/D_0[1]$, Eq.\ (\ref{3.20}), for different values of the parameters of the system.  As in the previous figures, we assume that $\alpha_{22}=\alpha_{12}\equiv \alpha$ so that we reduce the parameter set of the problem to three quantities: $\{\alpha,\mu,\omega\}$.

\begin{figure}
\includegraphics[width=0.4 \columnwidth]{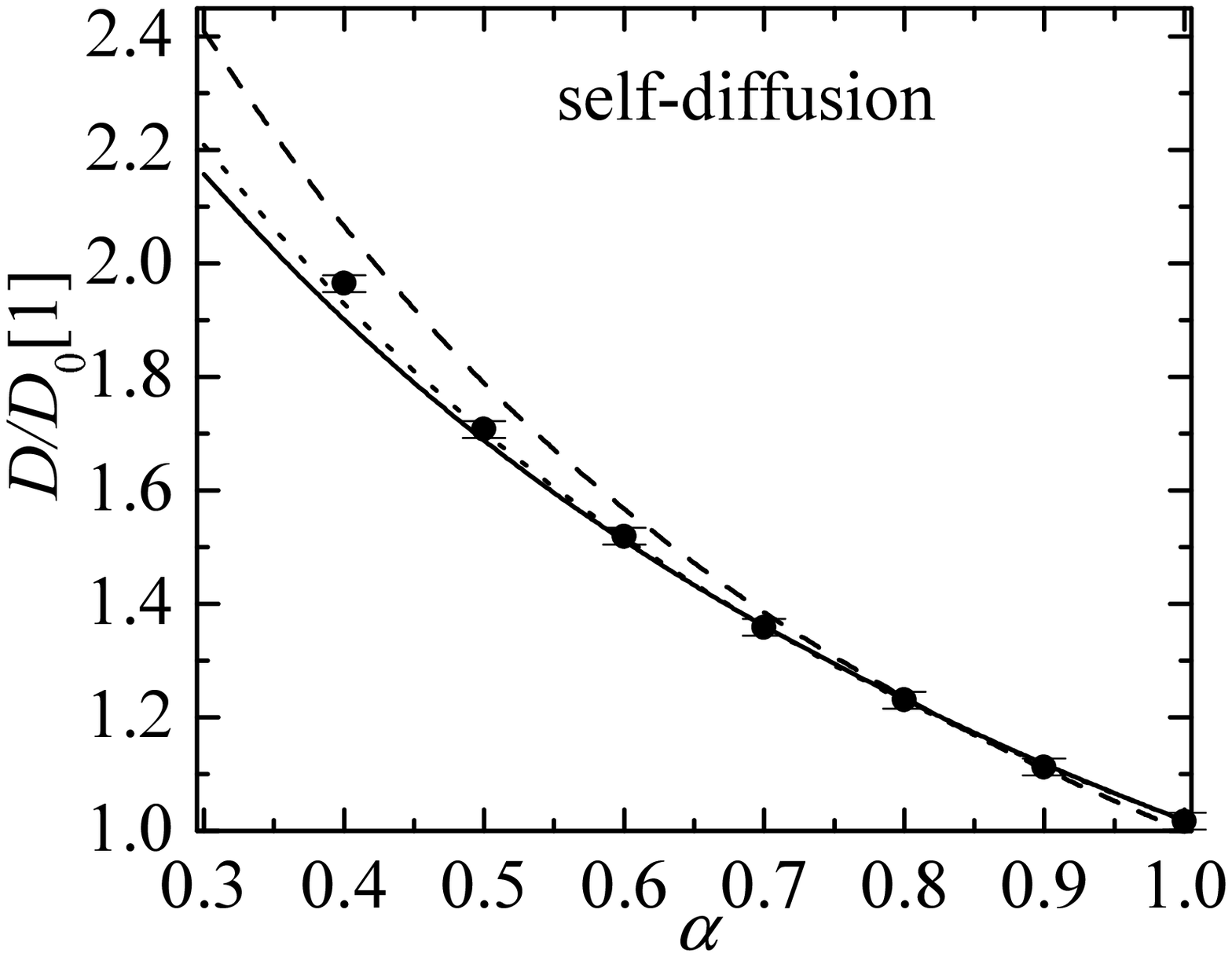}
\caption{Plot of the reduced self-diffusion coefficient $D/D_0[1]$ as a function of the coefficient of restitution $\alpha$ as given by the first Sonine approximation (dashed line), the second Sonine approximation (solid line), and Monte Carlo simulations (symbols). Here, $D_0[1]$ refers to the first Sonine approximation to the elastic value of the self-diffusion coefficient. The dotted line corresponds to the results obtained for the second Sonine approximation when one takes $c_1=c_2=0$. 
\label{fig4}}
\end{figure}

\begin{figure}
\includegraphics[width=0.4 \columnwidth]{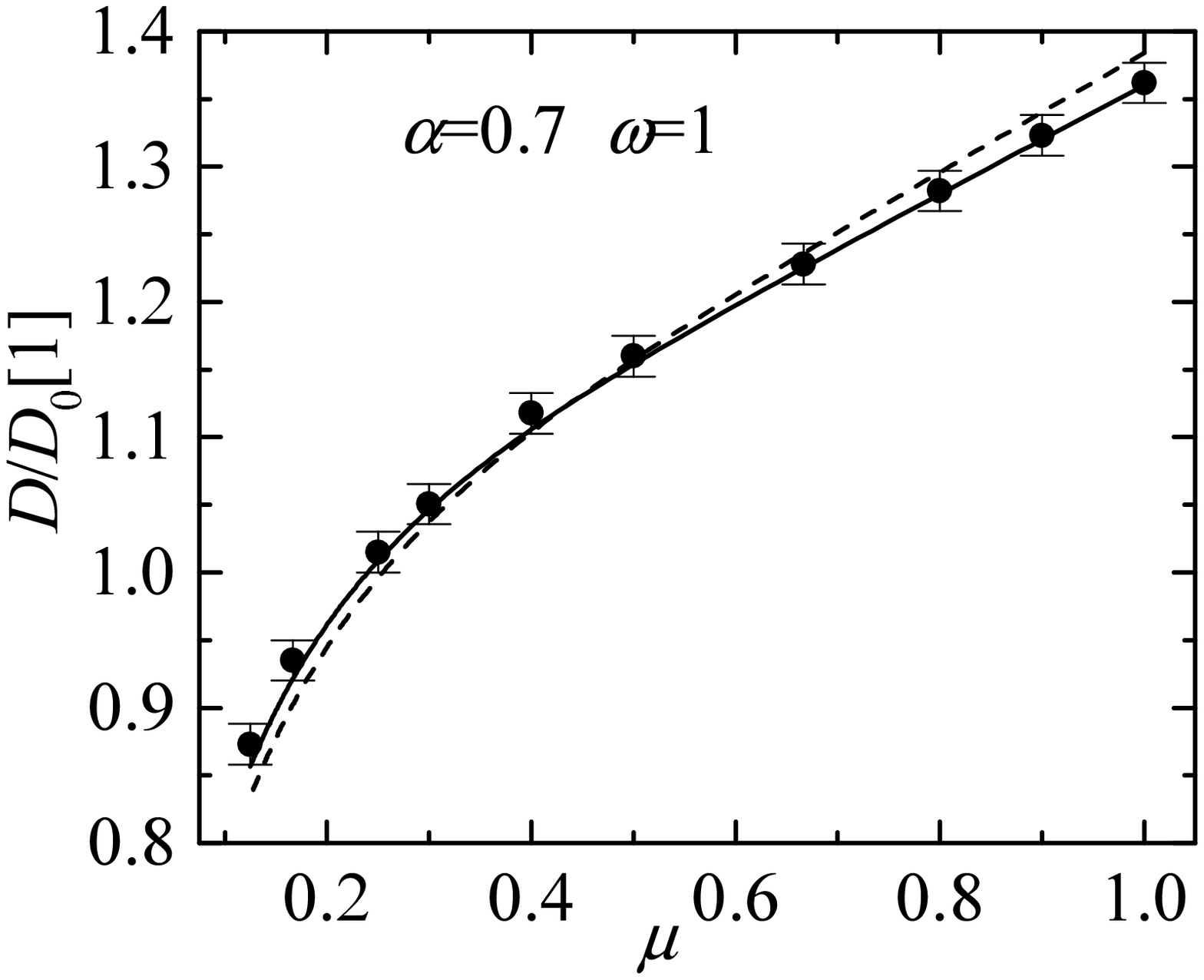}
\caption{Plot of the reduced diffusion coefficient $D/D_0[1]$ as a function of the mass ratio $\mu$ for $\alpha=0.7$ and $\omega=1$. The dashed line refers to the first Sonine approximation, the solid line corresponds to the second Sonine approximation while the symbols are the results obtained from Monte Carlo simulations. Here, $D_0[1]$ refers to the first Sonine approximation to the elastic value of the diffusion coefficient. 
\label{fig5}}
\end{figure}

\begin{figure}
\includegraphics[width=0.4 \columnwidth]{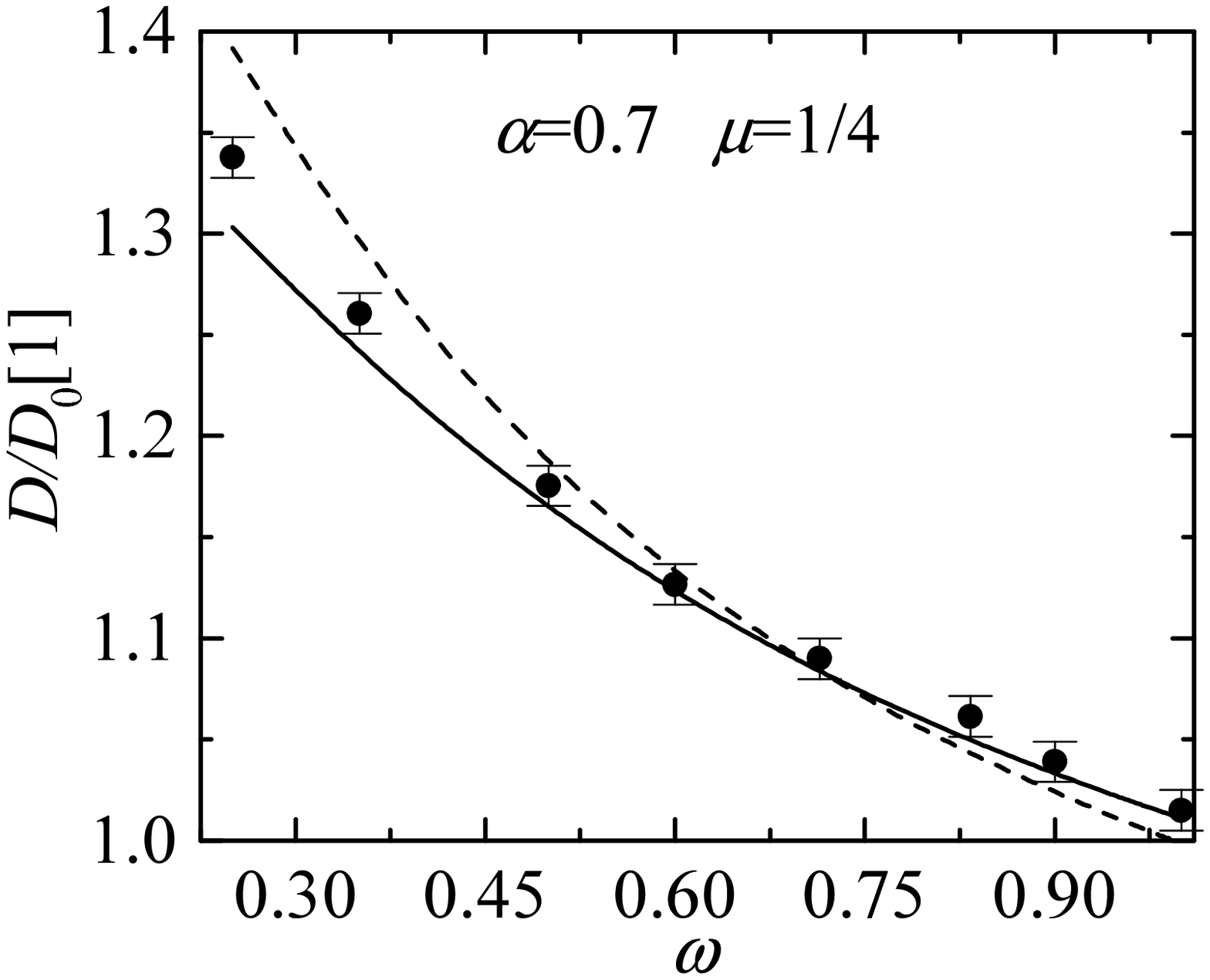}
\caption{Plot of the reduced diffusion coefficient $D/D_0[1]$ as a function of the size ratio $\omega$ for $\alpha=0.7$ and $\mu=1/4$. The dashed line refers to the first Sonine approximation, the solid line corresponds to the second Sonine approximation while the symbols are the results obtained from Monte Carlo simulations. Here, $D_0[1]$ refers to the first Sonine approximation to the elastic value of the diffusion coefficient.
\label{fig6}}
\end{figure}

Let us consider first the self-diffusion case ($\mu=\omega=1$). As said above, this special situation was studied by Brey {\em et al.} \cite{BMCG00} analytically (up to the first Sonine approximation) and by computer simulations. Figure \ref{fig4} shows the reduced diffusion coefficient $D(\alpha)/D_0[1]$ as a function of the coefficient of restitution $\alpha$ as given by the first Sonine approximation (dashed line), the second Sonine approximation (solid line) and Monte Carlo simulations (symbols). Although the agreement between the first Sonine approximation and simulation is quite good beyond the quasielastic limit (say for instance, $\alpha \geq 0.8$), discrepancies between both results increase as the value of $\alpha$ decreases. This tendency was also observed in the comparison made in Ref.\ [\onlinecite{BMCG00}]. We observe that the agreement between theory and simulation improves over the whole range of values of $\alpha$ considered when one takes the second Sonine approximation. For instance, for $\alpha=0.5$ the first and second Sonine approximations to $D$ differ by 4.7\% and 1.2\%, 
respectively, from the value measured in the simulation.  We have also included (dotted line) the result obtained for the second Sonine approximation to $D(\alpha)/D_0[1]$ if the distribution functions of the granular gas $f_2$ and impurities $f_1^{(0)}$ at zeroth-order were both approximated by Gaussians, i.e., when one formally puts $c_1=c_2=0$ in Eq.\ (\ref{3.19}). We observe that the influence of these Sonine contributions to the self-diffusion coefficient is negligible, except for quite large values of dissipation.

Consider now the situation in which impurities and particles of the gas can differ in size and mass. First, in Fig.\ \ref{fig5} we analyze the effect of the mass ratio on the accuracy of the Sonine approximations to the reduced diffusion coefficient $D(\alpha)/D_0[1]$. Here, we take  $\sigma_1=\sigma_2$ and $\alpha=0.7$. We find that, for fixed $\omega$ and $\alpha$, the second Sonine approximation $D[2]$ differs from the first Sonine approximation $D[1]$ as $\mu$ is varied. For the case considered in Fig.\ \ref{fig5}, we see that $D[2]>D[1]$ if $\mu \lesssim 0.45$ while $D[1]>D[2]$ otherwise. The comparison with simulation data shows again that  the theoretical predictions are improved when one considers the second Sonine approximation, showing an excellent agreement between $D[2]$ and Monte Carlo simulations in the range of values of $\mu$ explored. Similar conclusions are found with respect to the variations of $D(\alpha)/D_0[1]$ on the size ratio $\omega$ at fixed values of $\mu$ and $\alpha$, as shown in Fig.\ \ref{fig6} for $\alpha=0.7$ and $\mu=1/4$. Differences between both Sonine approximations are especially important for small values of $\omega$. In this region, the discrepancies between $D[2]$ and simulation data are slightly larger than the ones found before in Fig.\ \ref{fig5} for the mass ratio. This means that perhaps the most significant variations of the ratio $D[2]/D[1]$ occur when the size ratio $\omega$ is varied at fixed $\mu$.  

\begin{figure}
\includegraphics[width=0.4 \columnwidth]{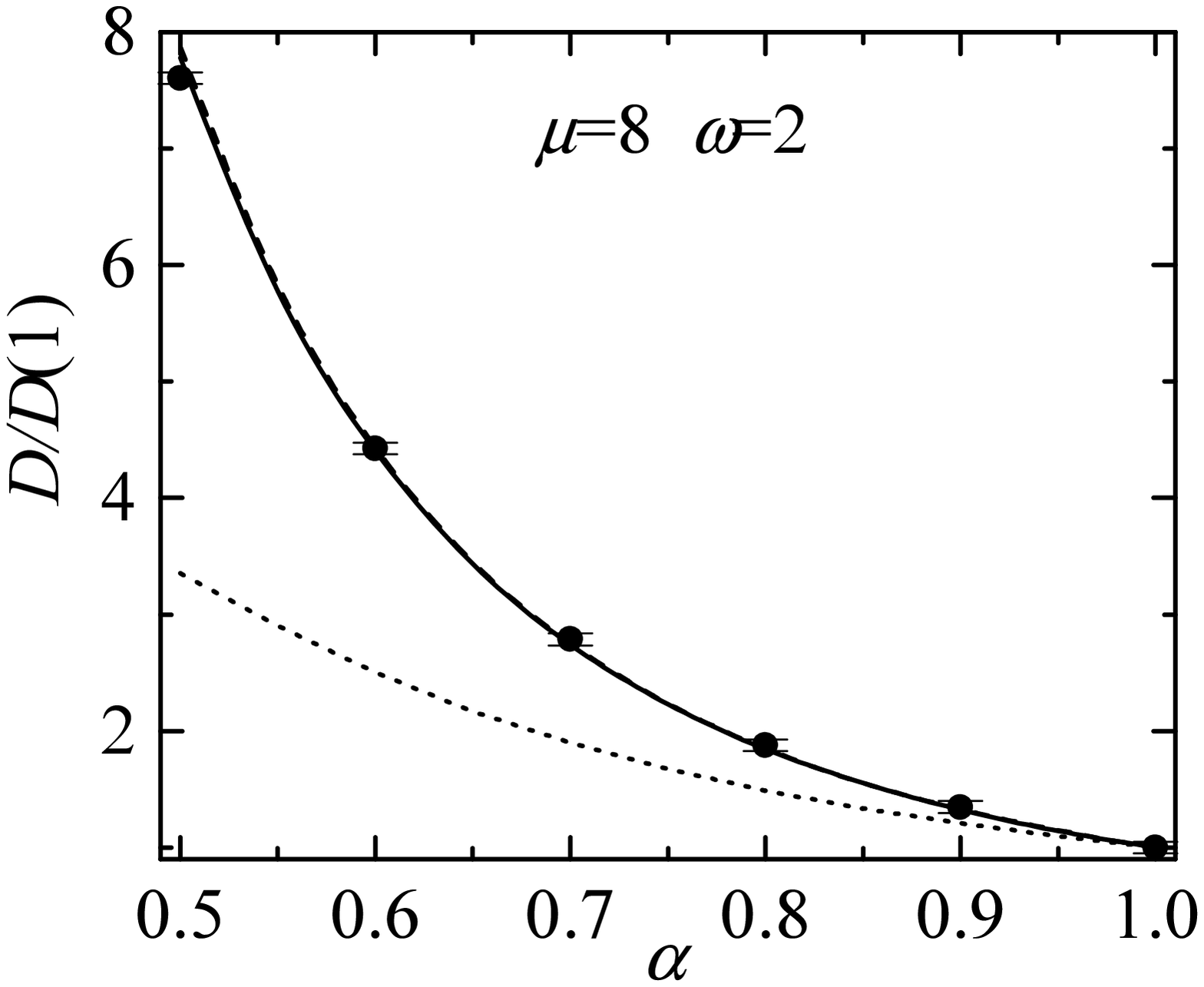}
\caption{Plot of the reduced diffusion coefficient $D(\alpha)/D(1)$ as a function of the coefficient of restitution $\alpha$ for $\mu=8$ and $\omega=2$. The dashed line refers to the first Sonine approximation, the solid line corresponds to the second Sonine approximation while the 
symbols are the results obtained from Monte Carlo simulations. The dotted line is the first Sonine approximation by assuming the equality of the partial temperatures $\gamma=T_1/T=1$. Here, $D(1)$ refers to the elastic value of the diffusion coefficient consistently obtained in each approximation. 
\label{fig7}}
\end{figure}

\begin{figure}
\includegraphics[width=0.4 \columnwidth]{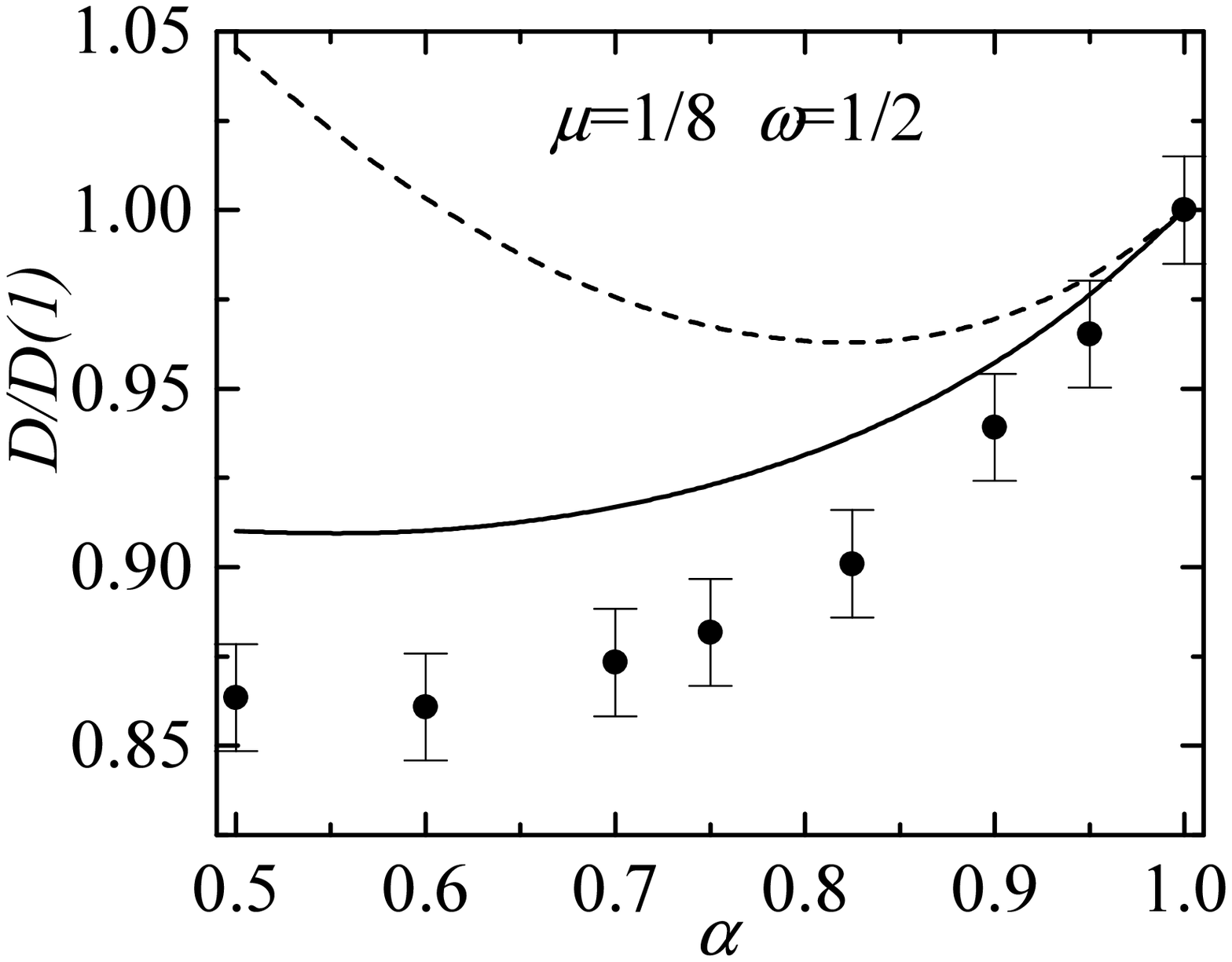}
\caption{Plot of the reduced diffusion coefficient $D(\alpha)/D(1)$ as a function of the coefficient of restitution $\alpha$ for $\mu=1/8$ and $\omega=1/2$. The dashed line refers to the first Sonine approximation, the solid line corresponds to the second Sonine approximation while the symbols are the results obtained from Monte Carlo simulations. Here, $D(1)$ refers to the elastic value of the diffusion coefficient consistently obtained in each approximation. 
\label{fig8}}
\end{figure}

Finally, we explore the influence of dissipation on the two first Sonine approximations at fixed values of $\mu$ and $\omega$. According to Fig.\ \ref{fig2}, it is apparent that the deviations of $\Lambda\equiv D[2]/D[1]$ from unity are important for small values of the mass ratio and/or the size ratio. In fact, for elastic mixtures ($\alpha=1$), $\Lambda \simeq 1.069$ for the case $\{\mu=1/8, \omega=1/2\}$, while $\Lambda \simeq 1.00049$ for the symmetric case $\{\mu=8, \omega=2\}$. This clearly shows that in the elastic case the second Sonine approximation significantly differs from the first one for small values of $\mu$ and $\omega$ (Lorentz gas limit), while both approximations practically coincide when $\mu$ and $\omega$ are much larger than 1 (Rayleigh gas limit). In the former case, where the tracer diffusion coefficient can be obtained exactly, the first and second Sonine approximations to $D$ differ by 13\% and 5\%, respectively, from the exact value of the diffusion coefficient \cite{MC84}. 
Here, we want to analyze the influence of  the inelasticity on the trends already observed in  ordinary fluids. In order to assess this effect, it is convenient to reduce the diffusion coefficient  $D(\alpha)$ with respect to its elastic value $D(1)$ consistently obtained in each approximation.  Thus, $D(1)\to D_0[1]$, Eq.\ (\ref{3.22}), when one considers the first Sonine approximation, while $D(1)\to D_0[2]$, Eq.\ (\ref{3.23}), when one considers the second Sonine approximation. Simulation data will be also reduced with respect to the elastic value measured in the simulation. Figure \ref{fig7} shows $D(\alpha)/D(1)$ for $\mu=8$ and $\omega=2$ while Fig.\ \ref{fig8} shows $D(\alpha)/D(1)$ for $\mu=1/8$ and $\omega=1/2$. The first finding is that the conclusions obtained in the elastic case on the convergence of Sonine polynomial expansion are kept at a qualitative level for granular gases: the Sonine polynomial expansion exhibits a poor convergence for sufficiently small values of the mass ratio $\mu$ and/or the size ratio $\omega$, while this convergence improves significantly as $\mu$ and/or $\omega$ increases. As a matter of fact, Fig.\ \ref{fig7} shows that both Sonine approximations are practically indistinguishable and present an excellent agreement with simulation data. This is not so in the case of Fig.\ \ref{fig8}, where only the second Sonine approximation to $D$ exhibits a qualitative good agreement with simulation results. As in the elastic case \cite{MC84}, one perhaps would have to consider the third Sonine approximation to provide an accurate estimate for the diffusion coefficient.

\section{Discussion}
\label{sec5}

 In this paper we have analyzed the diffusion of impurities in a dilute granular gas undergoing homogenous cooling state (HCS). This is the simplest example of transport in a {\em multicomponent} granular gas, since the gas is in a homogeneous state while diffusion appears by the presence of a weak concentration gradient, which is the only gradient present in the system. Here, the diffusion coefficient $D$ has been computed by combining two complementary approaches: a Chapman--Enskog solution to the Boltzmann--Lorentz equation and numerical solutions of the same equation by means of the direct simulation Monte Carlo (DSMC) method. As in the elastic case, the diffusion coefficient $D$ is given in terms of the solution of an integral equation, Eq.\ (\ref{3.11}). A practical evaluation of $D$ is possible by using a Sonine polynomial expansion and approximate results are {\em not} limited to weak inelasticity. Here, we have determined $D$ in the first (one Sonine polynomial) and second (two Sonine polynomials) Sonine approximation as a function of the temperature and the different mechanical parameters of the system, namely, the (constant) coefficients of restitution for the impurity--gas and gas--gas collisions, the masses and the particle sizes.  Our study complements and extends previous works on diffusion in undriven granular gases in the cases of self-diffusion \cite{BMCG00,LBD02} and Brownian motion \cite{BDS99,BMGD99}, and provides an explicit expression of the coefficient $D$ beyond the first Sonine approximation \cite{DG01,GD02,DBL02}.

Comparison between simulation data and theory shows that in general the second Sonine approximation $D[2]$ improves significantly the predictions of the first Sonine approximation $D[1]$, especially for values of the mass ratio $\mu$ and/or the size ratio $\omega$ smaller than 1. For this range of values of $\mu$ and $\omega$ (see, for instance, Fig.\ \ref{fig8}) one should consider higher-order polynomial terms to get a quantitative good agreement with Monte Carlo simulations. However, the convergence of the Sonine polynomial expansion (see, for instance, Fig.\ \ref{fig7}) improves with increasing values of $\mu$ and $\omega$ and the first Sonine correction seems to be quite close to the exact value. These trends are quite similar to those previously found for ordinary fluid mixtures \cite{MC84}.

Apart from extreme mass or size ratios, our results show again that the accuracy of the Sonine polynomial solution for granular systems is similar to that for elastic collisions. Possible discrepancies between theory and simulation can be removed by retaining higher-order Sonine corrections. On the other hand, as in all previous studies \cite{MG03,BMCG00,BMGD99,LBD02}, the good agreement found here over quite a wide range of values of the coefficient of restitution and mass and size ratios is an additional evidence of the validity of the hydrodynamic description to analyze some states of granular fluids.

One of the main features of our theory is that it incorporates the effect of energy nonequipartition on diffusion. This new effect had already been considered in some previous studies \cite{BDS99,MG03,DG01,GD02,DBL02}. To assess this effect  on the diffusion coefficient $D$, in Fig.\ \ref{fig7} we include for comparison the result for $D[1]$ with $\mu=8$ and $\omega=2$ that would be obtained if the differences in the partial temperatures were neglected ($T_1=T$). Clearly, inclusion of this effect makes a significant difference over the whole range of dissipation considered (for instance, the actual value of $\gamma$ is $\gamma\simeq 2.35$ for $\alpha=0.8$).

\begin{figure}
\includegraphics[width=0.4 \columnwidth]{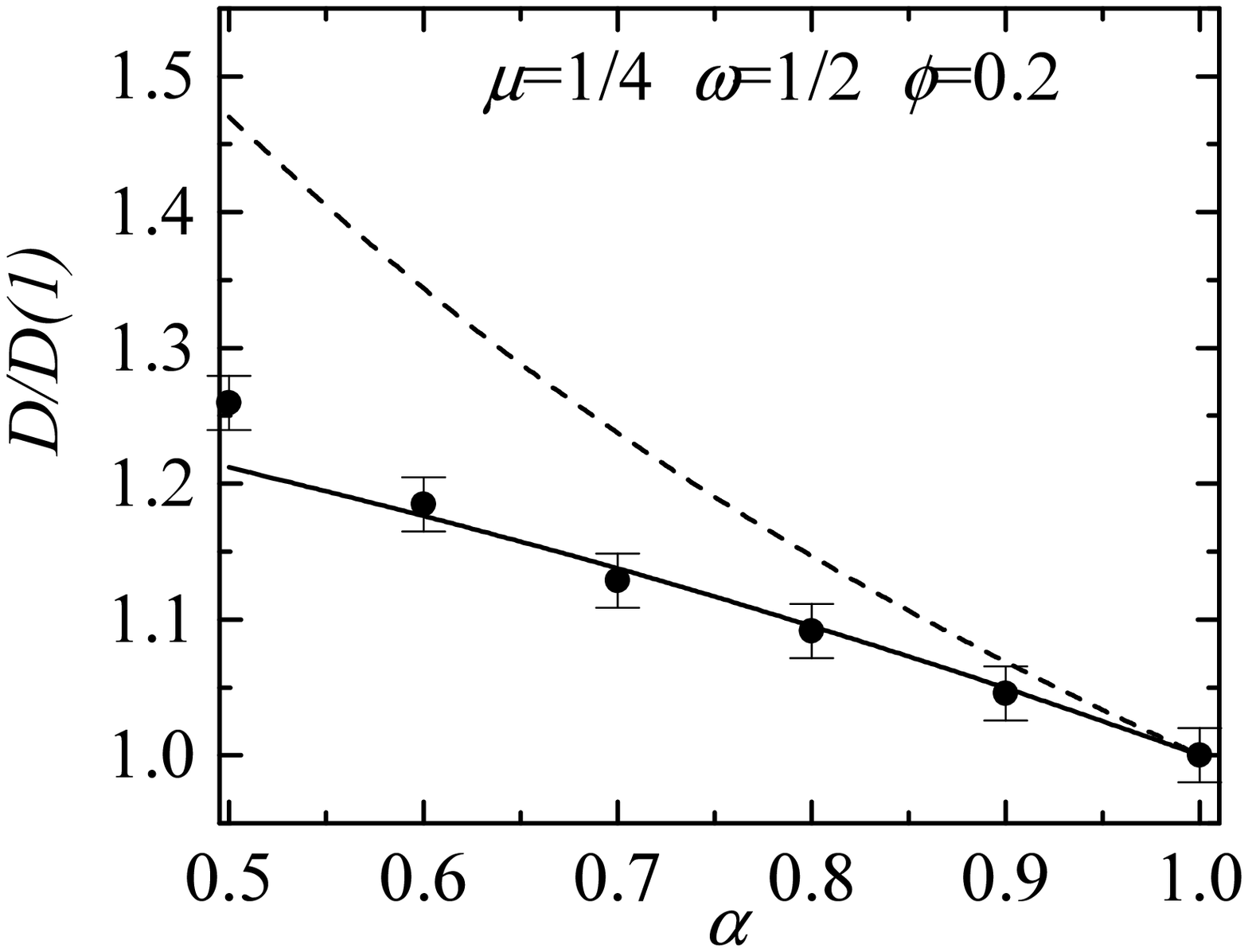}
\caption{Plot of the reduced diffusion coefficient $D(\alpha)/D(1)$ as a function of the coefficient of restitution $\alpha$ for $\mu=1/4$, $\omega=1/2$, and $\phi=0.2$. The dashed line refers to the first Sonine approximation, the solid line corresponds to the second Sonine approximation, while the symbols are the results obtained from Monte Carlo simulations. Here, $D(1)$ refers to the elastic value of the diffusion coefficient consistently obtained in each approximation. 
\label{fig9}}
\end{figure}

The results derived here are restricted to the low-density regime, small gradient in the concentration gradient, and to a system in the HCS. The HCS is known to be unstable under long wavelength perturbations or fluctuations \cite{GZ93}, with the critical wavelength $\lambda_c=2 \pi \sqrt{\eta/2 \zeta}$, where $\eta$ is the shear viscosity of the granular gas. Thus, our expression for the diffusion coefficient is only meaningful if the time for instability is longer than the few collision times required to reach the hydrodynamic regime. At low density this is clearly the case, even at strong dissipation, as demonstrated by both Monte Carlo and molecular dynamics simulations \cite{BMCG00}.  With respect to the extension to dense gases, the Enskog equation provides a useful generalization of the Boltzmann equation to higher densities for a gas of hard spheres. Applied to the diffusion of impurities in a granular fluid in the HCS, the only difference between the Boltzmann equations (\ref{2.13}) and (\ref{3.1}) and their corresponding Enskog counterparts is the presence of the factors $\chi_{22}$ in (\ref{2.13}) and $\chi_{12}$ in (\ref{3.1}), which are the configurational pair correlation functions for the fluid--fluid and impurity--fluid pairs at contact, respectively. A good approximation for these functions is provided by the extended Carnahan-Starling form \cite{CS}:
\begin{equation}
\label{5.1}
\chi_{22}=\frac{1-\frac{1}{2}\phi}{\left(1-\phi\right)^{3}},
\end{equation}
\begin{equation}
\label{5.2}
\chi_{12}=\frac{1}{1-\phi}+\frac{3}{2}\frac{\phi}{\left(1-\phi\right)^{2}}\frac{\sigma_1}{\sigma_{12}}+
\frac{1}{2}\frac{\phi^2}{\left(1-\phi\right)^{3}}\left(\frac{\sigma_1}{\sigma_{12}}\right)^2,
\end{equation}
where $\phi=(\pi/6)n_2\sigma_2^3$ is the solid volume fraction.  According to Eqs.\ (\ref{5.1}) and (\ref{5.2}), the correlation functions have the same density dependence only when the size ratio is equal to 1. Given that the fluid is in a homogenous state, it follows that $\chi_{22}$ and $\chi_{12}$ are uniform. Thus, it is evident that, when properly scaled, the solutions obtained here for the Boltzmann and the Boltzmann--Lorentz equation can be directly translated to the Enskog equation.  To illustrate the influence of the density on the predictions previously made for a dilute gas,  in Fig.\ \ref{fig9} we plot $D/D(1)$ as a function of the coefficient of restitution for $\mu=\frac{1}{4}$, $\omega=\frac{1}{2}$, and a solid volume fraction $\phi=0.2$. As for a low-density gas, the second Sonine approximation clearly improves the results obtained from the first Sonine approximation. We see that the second Sonine approximation compares very well with simulation data, except for quite large dissipation ($\alpha \simeq 0.5)$.

In summary, there is growing theoretical support for the validity of hydrodynamic transport processes in granular fluids. However, some care is needed in translating properties of normal fluids to those with inelastic collisions. As seen here, in the case of the diffusion coefficient, the mass transport is strongly influenced by dissipation and this dependence may be of some relevance  in sedimentation problems, for example.

\acknowledgments  
We acknowledge the partial support of the Ministerio de Ciencia y Tecnolog\'{\i}a (Spain) through Grant No. BFM2001-0718 (V. G.) and Grant No. ESP2003-02859 (J.M.M.).

\appendix
\section{Explicit expressions of $\zeta_1$ and $c_1$}
\label{appC} 
 
In this Appendix, we will quote the expressions of the partial cooling rate $\zeta_1$ and the coefficient $c_1$. These results were obtained in Ref.\ \cite{GD99} for arbitrary values of $x_1$. Here, we will display both expressions in the tracer limit $x_1\to 0$. The cooling rate $\zeta_1$ is defined by Eq.\ (\ref{2.16.1bis}).  By using the leading Sonine approximations (\ref{2.14}) and (\ref{2.18}) and neglecting nonlinear terms in $c_1$ and $c_2$, $\zeta_1$ can be written as \cite{GD99}
\begin{equation}
\label{c1}
\zeta_1=\lambda_{10}+\lambda_{11}c_1+\lambda_{12}c_2,
\end{equation}
where 
\begin{equation}
\label{c2}
\lambda_{10}=\frac{8}{3}\sqrt{\pi}n_2\sigma_{12}^2v_0\mu_{21}
\left(\frac{1+\theta}{\theta}\right)^{1/2}(1+\alpha_{12}) 
\left[1-\frac{\mu_{21}}{2}(1+\alpha_{12})(1+\theta)\right],
\end{equation} 
\begin{equation}
\label{c3}
\lambda_{11}=\frac{1}{12}\sqrt{\pi}n_2\sigma_{12}^2v_0\mu_{21}
\frac{(1+\theta)^{-3/2}}{\theta^{1/2}}(1+\alpha_{12}) 
\left[2(3+4\theta)-3\mu_{21}(1+\alpha_{12})(1+\theta)
\right],
\end{equation}
\begin{equation}
\label{c4}
\lambda_{12}=-\frac{1}{12}\sqrt{\pi}n_2\sigma_{12}^2v_0\mu_{21}
\left(\frac{1+\theta}{\theta}\right)^{-3/2}(1+\alpha_{12}) 
\left[2+3\mu_{21}(1+\alpha_{12})(1+\theta)
\right].
\end{equation}
Here, $\theta=m_1T/m_2T_1=\mu/\gamma$ with $\mu\equiv m_1/m_2$ and $\gamma\equiv T_1/T$.

The coefficient $c_1$ is defined by 
\begin{equation}
\label{c5}
c_1=\frac{8}{15}\left(\frac{m_1^2}{4n_1T_1^2}\int\, d{\bf 
v}v^4f_1^{(0)}-\frac{15}{4}\right).
\end{equation}
This coefficient is detemined by substitution of Eqs.\ (\ref{2.14}) and (\ref{2.18}) into the Boltzmann-Lorentz equation (\ref{2.16}), multiplying that equation by $v^4$, and integrating over the velocity. When only linear terms in $c_1$ and $c_2$ are retained, the result is found to be \cite{GD99}
\begin{equation}
\label{c6}
-\frac{15}{2}\frac{\mu_{12}^2}{\theta^2}\zeta_1\left(1+\frac{c_1}{2}\right)=\Omega _{10}+\Omega _{11}c_1+\Omega _{12}c_2,
\end{equation}
where   
\begin{eqnarray}
\label{c7}
\Omega _{10} &=&2\sqrt{\pi}n_2\sigma_{12}^2v_0\mu_{12}^2\mu_{21} \frac{\left(
1+\theta \right) ^{-1/2}}{\theta^{5/2}}\left( 1+\alpha _{12}\right) 
\left[ -2\left( 6+5\theta\right) +\mu _{21}\left(1+\alpha
_{12}\right) \left( 1+\theta \right) \left( 14+5\theta \right)\right.\nonumber\\
& &\left. 
 -8\mu_{21}^{2}\left( 1+\alpha _{12}\right) ^{2}\left( 1+\theta \right) ^{2} 
+2\mu _{21}^{3}\left( 1+\alpha _{12}\right) ^{3}\left(
1+\theta \right) ^{3}\right],  
\end{eqnarray}
\begin{eqnarray}
\label{c8}
\Omega _{11} &= & \frac{\sqrt{\pi}}{8}n_2\sigma_{12}^2v_0\mu_{12}^2\mu_{21} 
\frac{\left(1+\theta \right) ^{-5/2}}{\theta^{5/2}}
\left( 1+\alpha _{12}\right) \left[ -2\left( 90+231\theta +184\theta
^{2}+40\theta ^{3}\right) \right.  \nonumber \\
&&+3\mu _{21}\left( 1+\alpha _{12}\right) \left( 1+\theta \right) \left(
70+117\theta +44\theta ^{2}\right) -24\mu _{21}^{2}\left( 1+\alpha _{12}\right)
^{2}\left( 1+\theta \right) ^{2}\left( 5+4\theta \right)  \nonumber \\
&& \left. +30\mu _{21}^{3}\left( 1+\alpha _{12}\right) ^{3}\left(
1+\theta \right) ^{3}\right]  \;,  
\end{eqnarray}
\begin{eqnarray}
\label{c9}
\Omega_{12} &=&\frac{\sqrt{\pi}}{8}n_2\sigma_{12}^2v_0\mu_{12}^2\mu_{21} 
\frac{\left(1+\theta \right) ^{-5/2}}{\theta^{1/2}}
\left( 1+\alpha
_{12}\right) \left[ 2\left( 2+5\theta \right) +3\mu _{21}\left( 1+\alpha
_{12}\right) \left( 1+\theta\right)\left(2+5\theta\right) \right.  \nonumber \\
& & \left. -24\mu _{21}^{2}\left( 1+\alpha _{12}\right) ^{2}\left(
1+\theta \right) ^{2}+30\mu _{21}^{3}\left( 1+\alpha _{12}\right) ^{3}\left(
1+\theta \right) ^{3}\right]  \;.  
\end{eqnarray}
The final expression of $c_1$ is obtained by substitution of Eq.\ (\ref{c1}) into Eq.\ (\ref{c6}) and neglecting nonlinear terms in $c_1$ and $c_2$. The result is
\begin{equation}
\label{c10}
c_1=-\frac{\lambda_{10}+\lambda_{12}c_2+\frac{2}{15}\mu_{12}^{-2}\theta^2\left(\Omega_{10}+\Omega_{12}c_2\right)}
{\frac{1}{2}\lambda_{10}+\lambda_{11}+\frac{2}{15}\mu_{12}^{-2}\theta^2\Omega_{11}}.
\end{equation}

Once the coefficient $c_1$ is given in terms of $\gamma$ and the 
parameters of the mixture, the temperature ratio $\gamma$ can be explicitly 
obtained by numerically solving the condition for equal cooling rates: 
\begin{equation}
\label{c11}
\lambda_{10}+\lambda_{11}c_1+\lambda_{12}c_2=\zeta_2,
\end{equation}
where $\zeta_2$ is given by Eq.\ (\ref{2.15bis}).

\section{First and second Sonine approximations}
\label{appA}

In this Appendix we determine the coefficients $a_{1}$ and $a_{2}$ in the first and second Sonine approximation. Substitution of Eq.\ (\ref{3.14}) into the integral equation (\ref{3.11}) gives  
\begin{eqnarray}
\label{a1}
-\zeta T\partial_T(a_{1}f_{1,M}{\bf v}+a_{2}f_{1,M}{\bf S}_1) 
 &-&a_{1}J_{12}[f_{1,M}{\bf v},f_2]
 - a_{2}J_{12}[f_{1,M}{\bf S}_1,f_2]
 \nonumber\\
 &=&
 -\left(\frac{\partial}{\partial x_1}f_1^{(0)}\right){\bf v}.
\end{eqnarray}
Next, we multiply Eq.\ (\ref{a1}) by $m_1{\bf v}$ and 
integrate over the velocity. The result is 
\begin{equation}
\label{a2}
(-\zeta T\partial_T +\nu_a)n_1T_1a_{1}+n_1T_1\nu_b a_{2}=-
n_2 T_1.
\end{equation}
Here, $\zeta=\zeta_2$ is given by Eq.\ (\ref{2.15bis}) and we have introduced the quantities
\begin{equation}
\label{a3}
\nu_a=-\frac{m_1}{3n_1T_1}\int\, d{\bf v}\, {\bf v}\cdot 
J_{12}[f_{1,M}{\bf v},f_2],
\end{equation}
\begin{equation}
\label{a4}
\nu_b=-\frac{m_1}{3n_1T_1}\int\, d{\bf v}\, {\bf v}\cdot 
J_{12}[f_{1,M}{\bf S}_1,f_2].
\end{equation}
From dimensional analysis $T_1a_{1}\sim T^{1/2}$ so the temperature derivative can be performed in Eq.\ (\ref{a2}) and the result is   
\begin{equation}
\label{a5}
(\nu_a-\case{1}{2}\zeta)a_{1}+\nu_b a_{2}=-x_1^{-1},
\end{equation}
where $x_1=n_1/n_2$. If only the first Sonine correction is retained (which means $a_{2}\to 0$), the solution to (\ref{a5}) is 
\begin{equation}
\label{a6}
a_{1}[1]=-\frac{x_1^{-1}}{\nu_a-\case{1}{2}\zeta}.
\end{equation}
Here, $a_{1}[1]$ denotes the first Sonine approximation to $a_{1}$. Equation (\ref{a6}) leads to the expression (\ref{3.20}) for the diffusion coefficient $D[1]$ when the second equality in (\ref{3.17}) is considered.

To close the problem, one multiplies Eq.\ (\ref{a2}) by 
${\bf S}_1({\bf v})$ and integrates over the velocity. Following identical mathematical steps as those made before, one gets
\begin{equation}
\label{a7}
(\nu_c-\zeta T_1^{-1})a_{1}+
(\nu_d-\case{3}{2}\zeta)a_{2}=-\frac{1}{2}\frac{c_1}{x_1T_1},
\end{equation} 
where $c_1$ is given by Eq.\ (\ref{c10}) and we have taken into account that $T_1^3a_{2}\sim T^{3/2}$. Moreover, we have introduced the quantities
\begin{equation}
\label{a9}
\nu_c=-\frac{2}{15}\frac{m_1}{n_1T_1^3}\int\, d{\bf v}\, {\bf 
S}_1\cdot J_{12}[f_{1,M}{\bf v},f_2],
\end{equation}
\begin{equation}
\label{a10}
\nu_d=-\frac{2}{15}\frac{m_1}{n_1T_1^3}\int\, d{\bf v}\, {\bf S}_1\cdot J_{12}[f_{1,M}{\bf S}_1,f_2].
\end{equation}

In reduced units and by using matrix notation, Eqs.\ (\ref{a5}) and (\ref{a7}) can be rewritten as 
\begin{equation}
\label{a11}
\left(
\begin{array}{cc}
\nu_a^*-\case{1}{2}\zeta^{*}& \nu_b^*\\
\nu_c^*-\zeta^{*}/\gamma& \nu_d^*-\case{3}{2}\zeta^{*}
\end{array}
\right)
\left(
\begin{array}{c}
a_{1}^*\\
a_{2}^*
\end{array}
\right)
=-
\left(
\begin{array}{c}
1\\
c_1/2\gamma
\end{array}
\right).
\end{equation}
Here, $\zeta^*=\zeta/\nu_0$, $\nu_a^*=\nu_a/\nu_0$, 
$\nu_b^*=\nu_b/T\nu_0$, $\nu_c^*=T\nu_c/\nu_0$, and $\nu_d^*=\nu_d/\nu_0$ with $\nu_0=n_2\sigma_2^2v_{0}$. Further, $a_{1}^*=x_1\nu_0a_{1}$ and $a_{2}^*=x_1T\nu_0a_{2}$. The solution to Eq.\ (\ref{a11}) provides the explicit expression of the second Sonine approximation $a_{1}^*[2]$ to $a_{1}^*$:
\begin{equation}
\label{a12}
a_{1}^*[2]=a_{1}^*[1]\frac{(\nu_a^*-\case{1}{2}\zeta^*)
[\nu_d^*-\case{3}{2}\zeta^*-(c_1/2\gamma)\nu_b^*]}
{(\nu_a^*-\case{1}{2}\zeta^*)(\nu_d^*-\case{3}{2}\zeta^*)-
\nu_b^*[\nu_c^*-(\zeta^*/\gamma)]}.
\end{equation}
Equation (\ref{a12}) yields directly the expression (\ref{3.19}) for the second Sonine approximation $D[2]$.

\section{Evaluation of the collision integrals}
\label{appB}

In this Appendix we evaluate the quantities $\nu_a$, $\nu_b$, $\nu_c$, and $\nu_d$ defined by the collision integrals (\ref{a3}), (\ref{a4}), (\ref{a9}), and (\ref{a10}), respectively. Three of them, $\nu_a$, $\nu_c$, and $\nu_d$, were already determined in Ref.\ \cite{GD02} for arbitrary composition. For the sake of completeness, we display now their explicit expressions in the tracer limit ($x_1\to 0$). In reduced units, they are given by   
\begin{equation}
\label{b1}
\nu_a^*=\frac{4}{3}\sqrt{\pi}\mu_{21}\left(\frac{\sigma_{12}}
{\sigma_2}\right)^2(1+\alpha_{12})\left(\frac{1+\theta}{\theta}\right)^{1/2}
\left[1-\frac{c_2}{16}\left(\frac{\theta}{1+\theta}\right)^{2}\right],
\end{equation}
\begin{equation}
\label{b2}
\nu_c^*=\frac{4}{15}\sqrt{\pi}\frac{\mu_{21}^2}{\mu_{12}}
\left(\frac{\sigma_{12}}{\sigma_2}\right)^2(1+\alpha_{12})
\left(\frac{\theta^3}
{1+\theta}\right)^{1/2} A_c,
\end{equation} 
\begin{equation}
\label{b3}
\nu_d^*=\frac{2}{15}\sqrt{\pi}\mu_{21}
\left(\frac{\sigma_{12}}{\sigma_2}\right)^2(1+\alpha_{12})
\left(\frac{\theta}{1+\theta}\right)^{3/2} 
\left(A_d-5\frac{1+\theta}{\theta}A_c\right), 
\end{equation}  
where 
\begin{eqnarray}
\label{b4}
A_c&=&5(1+2\beta)+\mu_{21}(1+\theta)\left[5(1-\alpha_{12})-2(7\alpha_{12}-11)
\beta \theta^{-1}\right]+18\beta^2\theta^{-1} \nonumber\\
& & +2\mu_{21}^2(2\alpha_{12}^2-3\alpha_{12}+4)\theta^{-1}(1+\theta)^2-
5\theta^{-1}(1+\theta)\nonumber\\
& & +\frac{c_2}{16}\frac{\theta}{(1+\theta)^2}\left\{3\theta^2\mu_{21}(1+
\alpha_{12})\left[4\mu_{21}(1+\alpha_{12})-5\right]+\theta\left[
2\mu_{12}\left(7\mu_{21}(1+\alpha_{12})-5\right)\right.\right.
\nonumber\\
& & \left. 
+\mu_{21}\left(-5(9+7\alpha_{12})+\mu_{21}(38+62\alpha_{12}+24\alpha_{12}^2)
\right)\right]-15+54\mu_{12}^2-20\mu_{21}(3+\alpha_{12})\nonumber\\
& & \left. +2\mu_{21}^2(40+19\alpha_{12}+6\alpha_{12}^2)+2\mu_{12}\left[
\mu_{21}(61+7\alpha_{12}-20)\right]\right\},
\end{eqnarray}
\begin{eqnarray}
\label{b5}
A_d&=& 2\mu_{21}^2\left(\frac{1+\theta}{\theta}\right)^2
\left(2\alpha_{12}^{2}-3\alpha_{12}+4\right)
(8+5\theta)\nonumber\\
& & 
-\mu_{21}(1+\theta)\left[2\beta\theta^{-2}(8+5\theta)(7\alpha_{12}
-11)+2\theta^{-1}(29\alpha_{12}-37)-25(1-\alpha_{12})\right]
\nonumber\\
& & +18\beta^2\theta^{-2}(8+5\theta)+
2\beta\theta^{-1}(25+66\theta)+5\theta^{-1}
(6+11\theta)-5(1+\theta)\theta^{-2}(6+5\theta)
\nonumber\\
& & +\frac{c_2}{16}\left(1+\theta\right)^{-2}\left\{15\theta^3
\mu_{21}(1+\alpha_{12})(4\mu_{21}(1+\alpha_{12})-5)
\right.\nonumber\\
& & +2\left[45+540\mu_{12}^2+16
\mu_{21}(\alpha_{12}-36)
+4\mu_{21}^2(134+5\alpha_{12}+6\alpha_{12}^2)\right.
\nonumber\\
& & \left.
-4\mu_{12}(148+\mu_{21}(7\alpha_{12}-263))\right]+
\theta^2\left[-30-\mu_{21}(267+217\alpha_{12})
\right.\nonumber\\
& & \left.
+14\mu_{21}^2
(17+29\alpha_{12}+12\alpha_{12}^2)
+10\mu_{12}(7\mu_{21}(1+\alpha_{12})-5)\right]
\nonumber\\
& & 
+\theta\left[-315+270\mu_{12}^2-2\mu_{21}(55\alpha_{12}+57)
+\mu_{21}^2(440+326\alpha_{12}+156\alpha_{12}^2)\right.\nonumber\\
& & \left.\left.
+2\mu_{12}(-2+\mu_{21}(7\alpha_{12}+277))\right]\right\}.
\end{eqnarray}
In the above expressions, $\beta=\mu_{12}-\mu_{21}\theta$ and $c_2$ and $\zeta^*$ are given by Eqs.\  (\ref{2.15}) and (\ref{3.21}), respectively.

It only remains to evaluate $\nu_b$, which is defined by the collision integral (\ref{a4}). 
To simplify the integral, a useful identity for an arbitrary function $h({\bf v}_{1})$ is given by  
\begin{eqnarray} 
\label{b7}
\int d{\bf v}_{1}h({\bf v}_{1})J_{12}\left[{\bf v}_{1}|f_{1},f_{2}\right] 
&=&\sigma _{12}^{2}\int \,d{\bf v}_{1}\,\int \,d{\bf v}_{2}f_{1}({\bf 
v}_{1})f_{2}({\bf v}_{2})\nonumber\\
& & \times  \int 
d\widehat{\boldsymbol {\sigma}}\,\Theta (\widehat{\boldsymbol {\sigma}} \cdot {\bf g})(\widehat{\boldsymbol {\sigma}}\cdot {\bf g})\,
\left[ h({\bf v}_{1}'')-h({\bf v}_{1})\right],   
\end{eqnarray} 
with  
\begin{equation}
\label{b8} 
{\bf v}_{1}''={\bf v}_{1}-\mu _{21}(1+\alpha _{12})(  
\widehat{\boldsymbol {\sigma }}\cdot {\bf g})\widehat{\boldsymbol {\sigma}}\;, 
\end{equation} 
and ${\bf g}={\bf v}_1-{\bf v}_2$ is the relative velocity. Use of 
(\ref{b7}) in Eq.\ (\ref{a4}) gives
\begin{equation}
\label{b9}
\nu_b=\frac{\pi}{6}\frac{m_1}{n_1T_1}\sigma_{12}^2\mu_{21}(1+\alpha_{12})\int d{\bf v}_{1}\,\int d{\bf v}_{2}gf_{1,M}({\bf v}_{1})f_{2}({\bf v}_{2}) \left[{\bf S}_1({\bf v}_1)\cdot {\bf g}\right].
\end{equation}
Substitution of the distribution function $f_2$ from Eqs.\ (\ref{2.12}) and (\ref{2.14}) gives 
\begin{eqnarray}
\label{b10} 
\nu_{b} &=&\frac{T\nu_0}{3\pi^2}\mu_{12}\left(\frac{\sigma_{12}}
{\sigma_2}\right)^2(1+\alpha_{12})\theta^{3/2}
\int d{\bf v}_{1}^*\,\int d{\bf v}_{2}^*e^{-\theta v_1^{*2}-v_2^{*2}}g^*\nonumber\\
& & \times \left[1+\frac{c_2}{4}
\left(v^{*4}-5v^{*2}+\frac{15}{4}\right)\right]\left(\theta v_1^{*2}-\case{5}{2}\right)(g^*\cdot {\bf v}_1^*),
\end{eqnarray} 
where ${\bf v}^*={\bf v}/v_0$ and ${\bf g}^*={\bf g}/v_0$.  The integrals appearing in (\ref{b10}) can be evaluated by the change of variables
\begin{equation}
\label{b11} 
{\bf x}={\bf v}_{1}^{\ast}-{\bf v}_{2}^{\ast},\quad {\bf y}=\theta  
{\bf v}_{1}^{\ast}+{\bf v}_{2}^{\ast},   
\end{equation} 
with the Jacobian $\left(1+\theta\right)^{-3}$.  The integrals can be easily performed and the final expression for the dimensionless quantity $\nu_b^*=\nu_b/T\nu_0$ is 
\begin{equation}
\label{b12}
\nu_b^*=\frac{2}{3}\sqrt{\pi}\mu_{12}\left(\frac{\sigma_{12}}
{\sigma_2}\right)^2(1+\alpha_{12})\left[\theta^3(1+\theta)\right]^{-1/2}
\left[1+\frac{3}{16}c_2\left(\frac{\theta}{1+\theta}\right)^{2}\right].
\end{equation}

\end{document}